\documentclass[aps,prd,twocolumn,superscriptaddress,showpacs]{revtex4}
\usepackage[dvips]{graphicx}
\usepackage{amsmath,amssymb,times}

\newcommand{\bequ}{\begin{equation}}
\newcommand{\eequ}{\end{equation}}
\newcommand{\bea}{\begin{eqnarray}}
\newcommand{\eea}{\end{eqnarray}}

\DeclareSymbolFont{boldletters}{OML}{cmm} {b}{it}
\DeclareSymbolFontAlphabet{\mathbit}{boldletters}
\DeclareMathSymbol{\alpha}{\mathalpha}{letters}{"0B}
\DeclareMathSymbol{\beta}{\mathalpha}{letters}{"0C}
\DeclareMathSymbol{\gamma}{\mathalpha}{letters}{"0D}
\DeclareMathSymbol{\delta}{\mathalpha}{letters}{"0E}
\DeclareMathSymbol{\epsilon}{\mathalpha}{letters}{"0F}
\DeclareMathSymbol{\zeta}{\mathalpha}{letters}{"10}
\DeclareMathSymbol{\eta}{\mathalpha}{letters}{"11}
\DeclareMathSymbol{\theta}{\mathalpha}{letters}{"12}
\DeclareMathSymbol{\iota}{\mathalpha}{letters}{"13}
\DeclareMathSymbol{\kappa}{\mathalpha}{letters}{"14}
\DeclareMathSymbol{\lambda}{\mathalpha}{letters}{"15}
\DeclareMathSymbol{\mu}{\mathalpha}{letters}{"16}
\DeclareMathSymbol{\nu}{\mathalpha}{letters}{"17}
\DeclareMathSymbol{\xi}{\mathalpha}{letters}{"18}
\DeclareMathSymbol{\pi}{\mathalpha}{letters}{"19}
\DeclareMathSymbol{\rho}{\mathalpha}{letters}{"1A}
\DeclareMathSymbol{\sigma}{\mathalpha}{letters}{"1B}
\DeclareMathSymbol{\tau}{\mathalpha}{letters}{"1C}
\DeclareMathSymbol{\upsilon}{\mathalpha}{letters}{"1D}
\DeclareMathSymbol{\phi}{\mathalpha}{letters}{"1E}
\DeclareMathSymbol{\chi}{\mathalpha}{letters}{"1F}
\DeclareMathSymbol{\psi}{\mathalpha}{letters}{"20}
\DeclareMathSymbol{\omega}{\mathalpha}{letters}{"21}
\DeclareMathSymbol{\varepsilon}{\mathalpha}{letters}{"22}
\DeclareMathSymbol{\vartheta}{\mathalpha}{letters}{"23}
\DeclareMathSymbol{\varpi}{\mathalpha}{letters}{"24}
\DeclareMathSymbol{\varrho}{\mathalpha}{letters}{"25}
\DeclareMathSymbol{\varsigma}{\mathalpha}{letters}{"26}
\DeclareMathSymbol{\varphi}{\mathalpha}{letters}{"27}
\DeclareMathSymbol{\Gamma}{\mathalpha}{letters}{"00}
\DeclareMathSymbol{\Delta}{\mathalpha}{letters}{"01}
\DeclareMathSymbol{\Theta}{\mathalpha}{letters}{"02}
\DeclareMathSymbol{\Lambda}{\mathalpha}{letters}{"03}
\DeclareMathSymbol{\Xi}{\mathalpha}{letters}{"04}
\DeclareMathSymbol{\Pi}{\mathalpha}{letters}{"05}
\DeclareMathSymbol{\Sigma}{\mathalpha}{letters}{"06}
\DeclareMathSymbol{\Upsilon}{\mathalpha}{letters}{"07}
\DeclareMathSymbol{\Phi}{\mathalpha}{letters}{"08}
\DeclareMathSymbol{\Psi}{\mathalpha}{letters}{"09}
\DeclareMathSymbol{\Omega}{\mathalpha}{letters}{"0A}

\begin{document}
\title{ 
A QCD-like theory with the ${Z}_{N_c}$ symmetry}

\author{Hiroaki Kouno}
\email[]{kounoh@cc.saga-u.ac.jp}
\affiliation{Department of Physics, Saga University,
             Saga 840-8502, Japan}

\author{Yuji Sakai}
\email[]{sakai@phys.kyushu-u.ac.jp}
\affiliation{Department of Physics, Graduate School of Sciences, Kyushu University,
             Fukuoka 812-8581, Japan}

\author{Takahiro Makiyama}
\affiliation{Department of Physics, Saga University,
             Saga 840-8502, Japan}

\author{Kouhei Tokunaga}
\affiliation{Department of Physics, Saga University,
             Saga 840-8502, Japan}

\author{Takahiro Sasaki}
\email[]{sasaki@phys.kyushu-u.ac.jp}
\affiliation{Department of Physics, Graduate School of Sciences, Kyushu University,
             Fukuoka 812-8581, Japan}

\author{Masanobu Yahiro}
\email[]{yahiro@phys.kyushu-u.ac.jp}
\affiliation{Department of Physics, Graduate School of Sciences, Kyushu University,
             Fukuoka 812-8581, Japan}

\date{\today}

\begin{abstract}
We propose a QCD-like theory with the $\mathbb{Z}_{N_c}$ symmetry. 
The flavor-dependent twisted boundary condition 
(TBC) is imposed on $N_c$ degenerate flavor quarks
in the SU($N_c$) gauge theory. 
The QCD-like theory is useful to understand the mechanism of color 
confinement. Dynamics of the QCD-like theory is studied 
by imposing the TBC on the Polyakov-loop extended Nambu-Jona-Lasinio (PNJL) model. 
The TBC model is applied to two- and three-color cases. 
The $\mathbb{Z}_{N_c}$ symmetry is preserved below some temperature $T_c$, 
but spontaneously broken above $T_c$. 
The color confinement below $T_c$ preserves the flavor symmetry. 
Above $T_c$, the flavor symmetry is broken, 
but the breaking is suppressed by 
the entanglement between the Polyakov loop and the chiral condensate. 
Particularly at low temperature, 
dynamics of the TBC model is similar to that of the PNJL model with 
the standard fermion boundary condition, indicating that 
the $\mathbb{Z}_{N_c}$ symmetry is a good approximate concept in the 
latter model even if the current quark mass is small. 
The present prediction can be tested in future by lattice QCD, 
since the QCD-like theory has no sign problem. 
\end{abstract}

\maketitle

\section{Introduction.} 
Understanding of the confinement mechanism is 
one of the most important subjects in hadron physics. 
According to Lattice QCD (LQCD), the system is in the confinement 
and chiral symmetry breaking phase at low temperature ($T$), but 
in the deconfinement and chiral symmetry restoration phase at high $T$. 
The confinement mechanism is, nevertheless, still unclear for several reasons. 
The main reason is that the exact symmetry is not found for the 
deconfinement transition and hence the order parameter is unknown. 
In the limit of zero current quark mass, the chiral condensate 
is an exact order parameter for the chiral restoration. 
In the limit of infinite current quark mass, 
on the contrary, the Polyakov loop becomes an exact order parameter 
for the deconfinement transition, since the $\mathbb{Z}_{N_c}$ symmetry 
is exact there. 
For the real world in which $u$ and $d$ quarks have small current 
quark masses, 
the chiral condensate is considered to be a good order parameter, but 
it is not clear whether the Polykov loop is a good order parameter.
In this paper, we approach this problem by proposing a QCD-like theory 
with the $\mathbb{Z}_{N_c}$ symmetry.

We start with the SU($N_c$) gauge theory with $N_f$ degenerate flavor quarks. 
The partition function $Z$ in Euclidean spacetime is described by  
\bea
Z=\int Dq D\bar{q} DA \exp[-S_0] 
\label{QCD-Z}
\eea
with the action
\bea
S_0=\int d^4x [\sum_{f}\bar{q}_f(\gamma_\nu D_\nu +m_f)q_f
+{1\over{4g^2}}{F_{\mu\nu}^{a}}^2], 
\label{QCD-S}
\eea
where $q_f$ is the quark field with flavor $f$ and current quark mass $m_f$, 
$D_\nu =\partial_\nu+iA_\nu$ is the covariant derivative 
with the gauge field $A_\nu$, $g$ is the gauge coupling and  $F_{\mu\nu}=\partial_\mu A_\nu -\partial_\nu A_\mu -i[A_\mu ,A_\nu ]=F_{\mu\nu}^aT^a$ with the SU($N_c$) generator $T^a$. 
The temporal boundary condition for quark is 
\bea
q_f(x, \beta=1/T )=-q_f(x, 0). 
\label{period-QCD}
\eea
The $\mathbb{Z}_{N_c}$ transformation
changes the ferimon boundary condition as~\cite{RW,Sakai}
\bea
q_f(x, \beta)=-\exp{(-i 2\pi k/{N_c})}q_f(x, 0)  
\label{period-QCD-Z}
\eea
for integer $k$, while the action $S_0$ keeps the original form \eqref{QCD-S} 
since the $\mathbb{Z}_{N_c}$ symmetry is 
the center symmetry of the gauge symmetry~\cite{RW}. 
The $\mathbb{Z}_{N_c}$ symmetry thus breaks down 
through the fermion boundary condition in QCD. 

\begin{figure}[htbp]
\begin{center}
\vspace{0.5cm}
\includegraphics[width=0.4\textwidth]{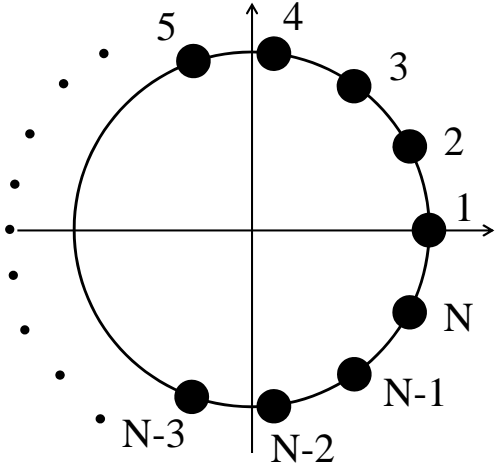}
\end{center}
\vspace{-10pt}
\caption{Twisted factors $e^{i\theta_f}$ ($f=1,2,\cdots, N$) 
on a unit circle in the complex plane for the case of $\theta_1 =0$. 
}
\label{theta_f}
\end{figure}

Now we consider the SU($N$) gauge theory with $N$ degenerate flavor 
quarks, i.e. $N \equiv N_f=N_c$, and assume 
the following twisted boundary conditions (TBC): 
\begin{eqnarray}
q_f(x, \beta )&=&-\exp{(i\theta_f)}q_f(x,0)
\nonumber\\
&\equiv& -\exp{[i(\theta_1+2\pi (f-1)/N)]}q_f(x, 0)
\label{period}
\end{eqnarray}
for flavors $f$ labeled by integers from 1 to $N$; 
see Fig.\ref{theta_f} for the twisted angles $\theta_f$. 
Here $\theta_1$ is an arbitrary real number in a range 
of $0 \le \theta_1 < 2\pi$. 
The action $S_0$ with the TBC is not QCD but a QCD-like theory. 
The QCD-like theory has the $\mathbb{Z}_{N_c}$ symmetry, 
i.e. invariant under the $\mathbb{Z}_{N_c}$ transformation. 
In fact, the $\mathbb{Z}_{N_c}$ transformation changes $f$ into $f-k$, 
but $f-k$ can be relabeled by $f$ since $S_0$ is invariant under 
the relabeling. 
The QCD-like theory with the $\mathbb{Z}_{N_c}$ symmetry 
is useful to understand the mechanism of color confinement. 

When the fermion field $q_f$ is transformed by 
\begin{eqnarray}
q_f \to \exp{(-i\theta_fT\tau )}q_f 
\label{transform_1}
\end{eqnarray}
for Euclidean time $\tau$, 
the action $S_0$ is changed into 
\bea
S(\theta_f)=\int d^4x [\sum_{f}\bar{q}_f
(\gamma_\nu D_\nu - \mu_f\gamma_4+m_f)q_f
+{1\over{4g^2}}F_{\mu\nu}^2] 
\notag \\
\label{QCD1}
\eea
with the imaginary quark number chemical potential $\mu_f=i T \theta_f$, 
while  
the TBC is transformed back to the standard one \eqref{period-QCD}. 
The action $S_0$ with the TBC is thus equivalent to 
the action $S(\theta_f)$ with the standard one \eqref{period-QCD}. 
In the limit of $T=0$, the action $S(\theta_f)$ 
tends to $S_0$ with \eqref{period-QCD} fixed. The QCD-like theory is thus 
identical with QCD at $T=0$ where 
the Polyakov loop $\Phi$ is zero. 
This indicates that in the QCD-like theory 
the $\mathbb{Z}_{N_c}$ symmetry is preserved up to some temperature $T_c$ 
and spontaneously broken above $T_c$. 
In the QCD-like theory, the flavor symmetry is explicitly broken by 
the flavor-dependent TBC. 
As shown later, the flavor-symmetry breaking is 
recovered below $T_c$. The breaking then becomes 
significant only above $T_c$.

In general, the QCD partition function $Z(T,\theta)$ with 
finite imaginary chemical potential 
$\theta$ has the Roberge-Weiss (RW) periodicity~\cite{RW}: 
$Z(T,\theta)=Z(T,\theta +2\pi k/N_c)$ for any integer $k$. 
The RW periodicity was confirmed 
by lattice QCD (LQCD)~\cite{FP,D'Elia,Chen34,Chen,D'Elia-iso,Cea,D'Elia-3, FP2010,Nagata,Takaishi} and the Holographic QCD~\cite{Aarts}.  
The RW periodicity means that $Z(T,\theta)$ is invariant 
under the extended $\mathbb{Z}_{N_c}$ transformation, i.e. 
the combination of the $\mathbb{Z}_{N_c}$ transformation and 
the parameter transformation $\theta \to \theta + 2 \pi k/N_c$. 
Actually, $Z(T,\theta)$ is transformed into 
$Z(T,\theta - 2 \pi k/N_c)$ by the $\mathbb{Z}_{N_c}$ transformation 
and $Z(T,\theta - 2 \pi k/N_c)$ is transformed back to $Z(T,\theta)$ by 
the parameter transformation. 
The QCD partition function thus has the extended $\mathbb{Z}_{N_c}$ symmetry, 
and dynamics of QCD at imaginary chemical potential 
is governed by the symmetry. 

The extended $\mathbb{Z}_{N_c}$ symmetry 
is not an internal symmetry, since the transformation includes the shift of 
external parameter $\theta$. In the QCD-like theory, the shift of $\theta$ 
is not necessary because of the TBC.
Thus the QCD-like theory possesses the $\mathbb{Z}_{N_c}$ symmetry 
as an internal symmetry, 
whereas QCD has the extended $\mathbb{Z}_{N_c}$ 
symmetry as an external symmetry. 
The Polyakov-loop extended Nambu-Jona-Lasinio (PNJL) model~\cite{Meisinger, Fukushima, Ratti, Rossner, Schaefer, Kashiwa1, Sakai, Sakai2, Kashiwa5, Matsumoto, Sasaki-T, Sakai5, Gatto, Sasaki-T_Nf3, Kahara,Sakai_hadron,Sakai_imiso,Brauner, Dumitru, Ghosh, Megias, Zhang, Mukherjee, Ciminale, Hansen, Sasaki, Costa, Fu, Abuki, Abuki2, McLerran_largeNc, Hell, Bhattacharyya, Fukushima3, Contrera,Kashiwa_MG, Kashiwa_NL_IM, Morita,Kouno} is a good model to understand 
QCD at finite imaginary chemical potential $\theta$ and hence 
the QCD-like theory, 
since the PNJL model possesses 
the extended $\mathbb{Z}_{N_c}$ symmetry in the standard fermion boundary 
condition \eqref{period-QCD}~\cite{Sakai}.

In this paper, we propose a QCD-like theory with 
the $\mathbb{Z}_{N_c}$ symmetry. 
The theory is constructed by imposing the TBC on 
the SU($N_c$) gauge theory with $N_c$ degenerate flavor quarks. 
Dynamics of the QCD-like theory is studied concretely by 
imposing the TBC on the PNJL model. 
Two cases of $N_c=N_f=2$ and 3 are mainly considered. 
In this paper, the PNJL model with the TBC is shortly 
called the TBC model, and the PNJL model with the standard boundary 
condition is named the standard-PNJL model. 
We first show that the $\mathbb{Z}_{N_c}$ symmetry is 
preserved below some temperature $T_c$, but spontaneously broken 
above $T_c$. 
The interplay between 
the $\mathbb{Z}_{N_c}$ symmetry breaking and the flavor symmetry breaking 
is investigated. 
Comparing the deconfinement transition in the TBC model with 
that in the standard-PNJL model, 
we show that the $\mathbb{Z}_{N_c}$ symmetry 
is a good approximate concept in the latter model, 
even if the current quark mass is small.  
The present prediction can be checked by LQCD in future, 
since LQCD with the TBC is free from the sign problem.

This paper is organized as follows. 
The case of $N_c=N_f=2$ is investigated in Sec. \ref{Sec:NC2} and 
that of $N_c=N_f=3$ is in Sec. \ref{Sec:NC3}. 
Two interesting extensions of the TBC model are shown 
in Sec. \ref{Two extensions of the TBC model}.  
Section \ref{Summary} is devoted to summary.

\section{Case of $N_c=2$}
\label{Sec:NC2}
\subsection{Formalism}
\label{Formalism-color-2}

The two-color and two-flavor PNJL Lagrangian~\cite{Brauner} 
in Euclidean spacetime is 
\begin{eqnarray}
\mathcal{L}&&=\sum_f
\bar{q}_f(\gamma_\nu D_\nu - \mu_f \gamma_4 + m_f )q_f
\nonumber\\
     &&-(1-\alpha )G_{\rm s} \sum_f \sum_{a=0}^3\left[({\bar q}_f\tau_a q_f)^2 
       +({\bar q}_fi\gamma_5 \tau_a q_f)^2\right]
\nonumber\\
     && + 4\alpha G_{\rm s}\left[\det_{ij}{\left
     (\bar{q}_{i}(1+\gamma_5)q_{\rm j}\right)}
               +\det_{ij}{\left(\bar{q}_{i}(1-\gamma_5)q_{j}\right)}\right]  
\nonumber\\
&&+{\cal U}(\Phi [A],{\Phi} [A]^*,T) 
\nonumber\\
\label{nc2_1}
\end{eqnarray}
with $D_\nu=\partial_\nu + iA_\nu=
\partial_\nu +i\delta_{\nu ,4}A_{4,a}{\tilde{\tau}_a\over{2}}$ 
for the gauge field $A^\nu_a$, 
where  the $\tau_a$ ($\tilde{\tau}_a$) for $a=1,2,3$ are 
the Pauli matrices in flavor (color) space and 
$\tau_0$ is the unit matrix in flavor space. 
In the NJL sector, $(1-\alpha)G_{\rm s}$ denotes coupling constants  
of scalar- and pseudoscalar-type four-quark interactions, whereas 
$\alpha G_{\rm s}$ is that 
of the Kobayashi-Maskawa-'t Hooft determinant interaction~\cite{KMK,tHooft}. 
Here $\alpha$ can vary from 0 to 1/2 for positive $G_{\rm s}$.  
The $U_{\rm A}(1)$ anomaly vanishes when $\alpha =0$. 
The Polyakov potential ${\cal U}$, defined in (\ref{nc2_3}), 
is a function of the Polyakov loop $\Phi$ and its Hermitian 
conjugate $\Phi^*$. 
The parameter $m_f$ ($\mu_f$) stands for the current quark mass 
(the chemical potential) for each flavor. Here we set $m_0 \equiv m_u=m_d$. 

In the PNJL model, the gauge field $A_\mu$ is treated as a homogeneous 
and static background field~\cite{Fukushima,Brauner}. 
In the case of $N_c=2$, the Polyakov-loop $\Phi$ and its conjugate $\Phi ^*$ 
are determined in Euclidean spacetime by 
\begin{align}
\Phi &= {1\over{2}}{\rm tr}_{\rm c}(L),
~~~~~\Phi^* ={1\over{2}}{\rm tr}_{\rm c}({\bar L}),
\label{Polyakov}
\end{align}
where $L  = \exp(i A_4/T)$ with $A_4=iA_0$. 
In the Polyakov-gauge, $A_4$ is diagonal in color space, i.e., 
$A_4/T={\rm diag}(\phi_1,\phi_2)$ for the $\phi_i$ 
satisfying $\phi_1+\phi_2=0$. 
This leads to 
\begin{eqnarray}
\Phi &=&{1\over{2}}(e^{i\phi_1}+e^{i\phi_2})
\notag\\
&=&{1\over{2}}(e^{i\phi_1}+e^{-i\phi_1})=\cos{(\phi_1)}, 
\notag\\
\Phi^* &=&{1\over{2}}(e^{-i\phi_1}+e^{-i\phi_2})
\notag\\
&=&{1\over{3}}(e^{-i\phi_1}+e^{i\phi_1})=\cos{(\phi_1)}=\Phi,  
\label{Polyakov_explict}
\end{eqnarray}
indicating that $\Phi$ is real. 
For the Polyakov-loop potential ${\cal U}$, we use 
\begin{eqnarray}
{\cal U}=-bT[24e^{-a/T}\Phi^2+\log{\left(1-\Phi^2\right)}] 
\label{nc2_3}
\end{eqnarray}
proposed in Ref.~\cite{Brauner}, where $a=858.1$~MeV and $b^{1/3}=210.5$~MeV.  
The Polyakov potential yields the second-order deconfinement phase transition 
at $T_c=270$~MeV in the pure gauge theory.

Now we consider the imaginary chemical potential $\mu_{f}=i\theta_fT$, where 
the twisted angles $\theta_f$ are real. 
Making the mean-field approximation (MFA) and the path integral 
over the quark fields in the PNJL partition function $Z_{\rm PNJL}$, 
one can obtain the thermodynamic potential (per unit volume) as 
\begin{align}
\Omega&=-T\ln(Z_{\rm PNJL})/V
\nonumber\\
&= -2 \sum_{f=u,d} \sum_{c=1,2}\int \frac{d^3 p}{(2\pi)^3}
   \Bigl[ E_f \nonumber\\
&        + \frac{1}{\beta}\ln~ [1 + e^{i\phi_c}e^{i\theta_f}e^{-\beta E_f}]
\notag\\
&        + \frac{1}{\beta}\ln~ [1 + e^{-i\phi_c}e^{-i\theta_f}e^{-\beta E_f}]
\Bigl]
\nonumber\\
&+ U(\sigma , a_0) +{\cal U}(\Phi,T), 
\label{nc2_2}
\end{align}
where $E_{f}^\pm({\bf p})=E_{f}({\bf p})\pm \mu_{\rm f}$ 
for $E_f({\bf p})=\sqrt{{\bf p}^2+{M_f}^2}$, 
\begin{eqnarray}
M_u&=&m_0-2G_s(\sigma +\zeta a_0), 
\label{nc2_mu}
\\
M_d&=&m_0-2G_s(\sigma -\zeta a_0), 
\label{nc2_md}
\\
U&=&G_s[\sigma^2+\zeta a_0^2], 
\label{nc_um}
\end{eqnarray}
$\zeta =1-2\alpha$, $\sigma =\langle \bar{u}u+\bar{d}d\rangle $ 
and  $a_0 =\langle \bar{u}u-\bar{d}d \rangle$. 
Here only the flavor-diagonal scalar condensates are taken.  
On the right-hand side of (\ref{nc2_2}) only the first term 
is regularized by the three-dimensional momentum cutoff 
$\Lambda$~\cite{Fukushima,Ratti}, since it diverges.

The variables, $X=(\Phi,{\Phi}^*, \sigma, a_0)$, are determined by 
the stationary conditions 
\bea
\partial \Omega/\partial X =0.
\label{eq:SC}
\label{condition}
\eea
Solutions $X(T,\theta_f)$ of the conditions do not give a 
global minimum of $\Omega$ necessarily, 
when the solutions  are inserted back to (\ref{nc2_2}). 
There is a possibility that they yield a local minimum or even a maximum. 
We have then checked that the solutions yield a global minimum

Following Ref.~\cite{Brauner}, 
we take the parameter set of $m_0=5.4$~MeV, $\Lambda =657$~MeV 
and $G_{\rm s}=7.23$~GeV$^2$ that yield 
$-\langle \bar{u}u\rangle ^{1/3}=218$~MeV, 
the pion decay constant $f_\pi =75.4$~MeV 
and the pion mass $m_\pi =140$~MeV at vacuum.

Taking the summation over color indices in \eqref{nc2_2} leads to  
\begin{eqnarray}
\Omega &=& -2\sum_{f=u,d}\int \frac{d^3{\rm p}}{(2\pi)^3}
         \Bigl[ 2 E_{f} \nonumber\\
       && + \frac{1}{\beta}
         \ln~ [1 + C_{2,1}({\bf p}) e^{i\theta_f}
         +C_{2,2}({\bf p})  e^{2i\theta_f}
         \nonumber\\
       && + \frac{1}{\beta} 
           \ln~ [1 + C_{2,1}({\bf p}) e^{-i\theta_f}
         +C_{2,2}({\bf p})  e^{-2i\theta_f}]
           \Bigl]\nonumber\\
       &&  + U(\sigma , a_0) +{\cal U}(\Phi ,T), 
\label{nc2_4} 
\end{eqnarray}
where 
\begin{eqnarray}
C_{2,1}({\bf p})&=&2\Phi e^{-\beta E_{f}}, 
\nonumber\\
C_{2,2}({\bf p})&=&e^{-2\beta E_{f}}.
\label{C_factor_2}
\end{eqnarray}
It is found from \eqref{C_factor_2} that 
$C_{2,1}=0$ and $C_{2,2} \ne 0$ when $\Phi=0$. 
The configuration means 
that two colored quarks are statistically in the same state. 
The configuration is thus 
realized as a result of the color confinement ($\Phi=0$). 
In other words, the color confinement can be defined by the configuration.

Making the $\mathbb{Z}_2$ transformation 
\begin{eqnarray}
\Phi \to e^{-ik\pi}\Phi
\label{nc_2_5}
\end{eqnarray}
in \eqref{nc2_4}, one can find that $\Omega$ has the RW periodicity,
\begin{eqnarray}
\Omega (\theta_f)=\Omega (\theta_f+k\pi). 
\label{RW_nc2}
\end{eqnarray}
Namely, $\Omega$ has the extended $\mathbb{Z}_2$ symmetry. 
The TBC corresponds to setting $\theta_u=\theta_1$ 
and $\theta_d=\theta_1 +\pi$ in $\Omega$. 
The $\mathbb{Z}_2$ symmetry with odd $k$ 
changes $(\theta_u,\theta_d)$ to 
$(\theta_d,\theta_u)$, but $(\theta_d,\theta_u)$ returns to 
$(\theta_u,\theta_d)$ by the relabeling of flavors. 
The TBC model thus has the $\mathbb{Z}_2$ symmetry as an internal symmetry 
in addition to the extended $\mathbb{Z}_2$ symmetry as an external symmetry.

In the color-confinement phase defined by $\Phi=0$, the thermodynamic 
potential $\Omega$ has only the 
configuration of $C_{2,1}=0$ and $C_{2,2} \ne 0$, as mentioned above. 
Components including $C_{2,2}$ 
in \eqref{nc2_4} have flavor dependence only through 
factors $e^{\pm 2i \theta_f}$, but the factors has no flavor dependence 
because of $(\theta_u,\theta_d)=(\theta_1,\theta_1 +\pi)$. 
Noting that the $M_f$ are determined by the stationary condition 
\eqref{condition} from the flavor-independent $\Omega$, one can see that 
the flavor symmetry is recovered by the color confinement.

\subsection{Numerical results}

Let us start with the standard-PNJL model, i.e., 
the PNJL model with no chemical potential. 
In this case, the Polykov loop $\Phi$ is an approximate 
order parameter of the color confinement, while the chiral condensate 
$\sigma$ is an approximate order parameter of the chiral transition. 
The flavor symmetry breaking is described by the isovector condensate $a_0$. 
We mainly consider the $U_{\rm A}(1)$ symmetric case by taking $\alpha =0$.

\begin{figure}[htbp]
\begin{center}
\hspace{-10pt}
\includegraphics[width=0.3\textwidth,angle=-90]{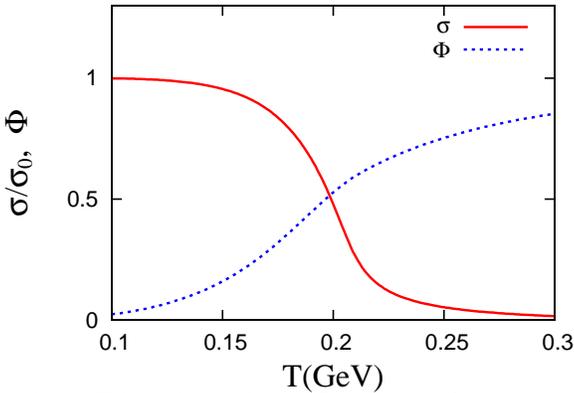}
\end{center}
\vspace{-10pt}
\caption{
$T$ dependence of $\sigma$ and $\Phi$ in the case of 
$\alpha =0$ and $(\theta_u,\theta_d)=(0,0)$. Here 
$\sigma$ is normalized by $\sigma_0 \equiv \sigma (T=0,\mu_f=0)$. 
}
\label{Fig_nc2_mu0}
\end{figure}

Figure~\ref{Fig_nc2_mu0} shows $T$ dependence of $\Phi$ and $\sigma$ 
calculated with the standard-PNJL model, where 
$\sigma$ is normalized by $\sigma_0\equiv \sigma (T=0,\mu_f=0)$. 
Both $\sigma$ and $\Phi$ are finite for any $T$, 
since there is no exact chiral and $\mathbb{Z}_2$ symmetry. 
As $T$ increases, $\sigma$ decreases gradually, 
while $\Phi$ increases smoothly. The chiral and deconfinement transitions 
are thus crossover. 
Here $a_0$ is zero at any $T$, since the flavor symmetry is not 
broken.

Now we consider the TBC model with the $\mathbb{Z}_2$ symmetry. 
The Polyakov loop $\Phi$ is an exact order parameter 
of the color confinement. 
When $\Phi \neq 0$, there are two $\mathbb{Z}_2$ vacua. 
The vacuum with positive $\Phi$ is taken in this paper.

\begin{figure}[htbp]
\begin{center}
\vspace{0.5cm}
\includegraphics[width=0.44\textwidth]{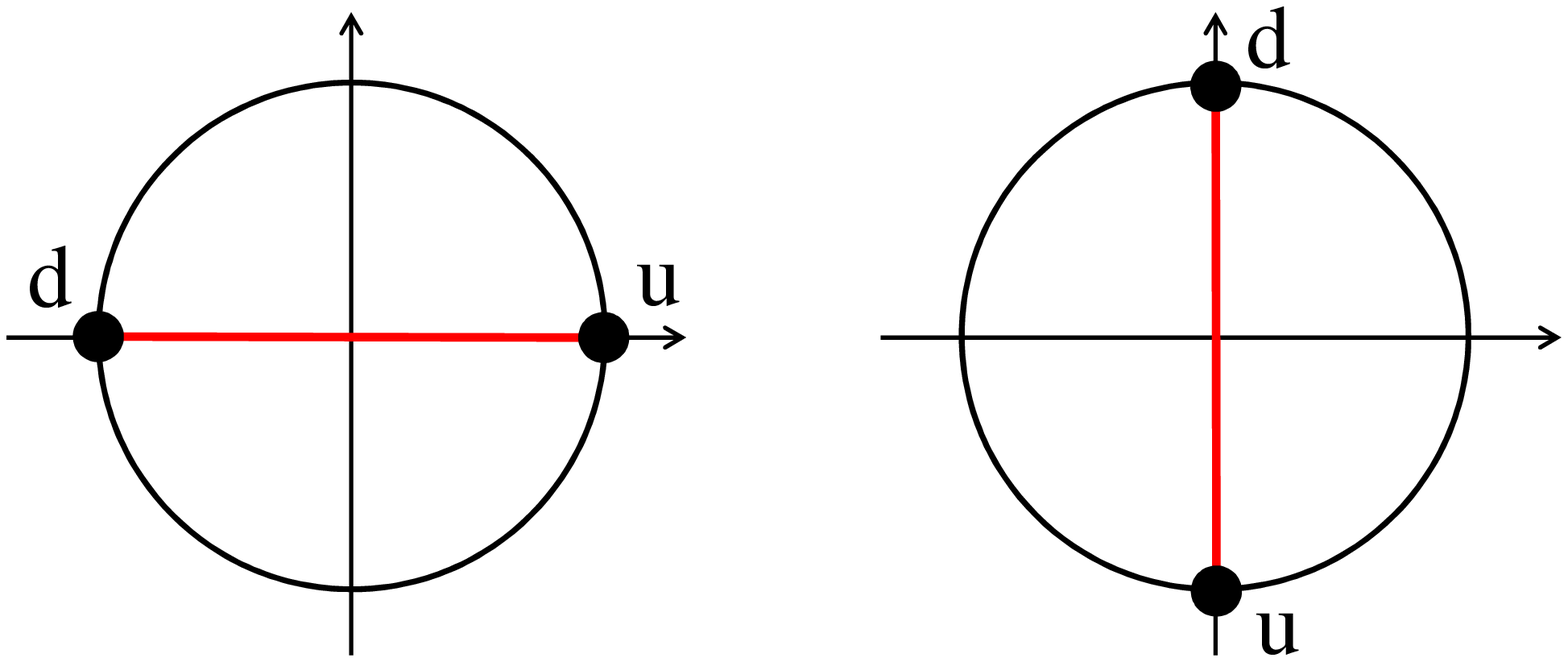}
\end{center}
\vspace{-10pt}
\caption{Twisted factors $e^{i\theta_f}$ on a unit circle 
in the complex plane 
for the case of $\theta_1 =0$ (left) and $\theta_1 =-{\pi\over{2}}$ (right). 
}
\label{theta_f_nc2}
\end{figure}

First we analyze the case 
\bea
(\theta_u,\theta_d)=(-\pi/2,\pi/2)
\label{case-2-1}
\eea
corresponding to $\theta_1 =-\pi/2$ in the TBC of \eqref{period}; 
see the right panel of Fig. \ref{theta_f_nc2} for the twisted angles.  
In this case, 
the flavor symmetry is not broken by the TBC, because 
\bea
\Omega(\theta_u,\theta_d)=\Omega(-\theta_u,-\theta_d)
=\Omega(\theta_d,\theta_u),
\eea
where the first and second equalities are 
obtained by the charge-conjugation and \eqref{case-2-1}, respectively.

\begin{figure}[htbp]
\begin{center}
\hspace{-10pt}
\includegraphics[width=0.295\textwidth,angle=-90]{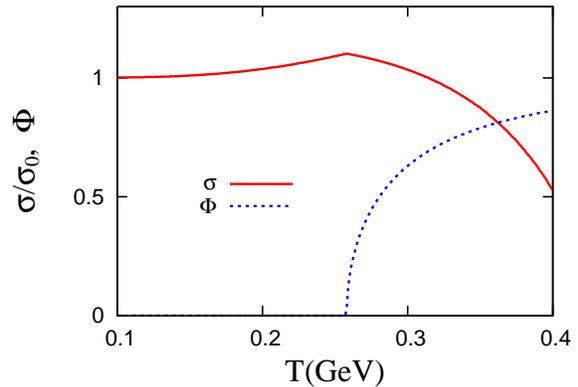}
\end{center}
\vspace{-10pt}
\caption{
$T$ dependence of $\sigma$ and $\Phi$ in the case of 
$\alpha =0$ and $(\theta_u,\theta_d)=(-\pi/2,\pi/2)$. 
$\sigma$ is normalized by $\sigma_0$. 
}
\label{Fig_nc2_order_z2b}
\end{figure}

Figure \ref{Fig_nc2_order_z2b} shows $\sigma$ and $\Phi$ 
as a function of $T$; 
note that $a_0$ is zero for any $T$ because of the flavor symmetry. 
The Polyakov loop $\Phi$ is zero up to 
$T \equiv T_c \approx 260$~MeV, but finite above $T_c$. 
The $\mathbb{Z}_2$ symmetry is thus preserved exactly below $T_c$, but 
spontaneously broken above $T_c$. 
The deconfinement phase transition is second-order, since 
$\Phi$ has no jump at $T=T_c$. 
Meanwhile, the chiral transition is crossover. 
There is no qualitative difference between the standard-PNJL model 
and the TBC model with $\theta_1=-\pi/2$ 
for the deconfinement and chiral transitions, 
although the order of the deconfinement transition becomes stronger 
by the exact $\mathbb{Z}_2$ symmetry.

\begin{figure}[htbp]
\begin{center}
\hspace{-10pt}
\includegraphics[width=0.3\textwidth,angle=-90]{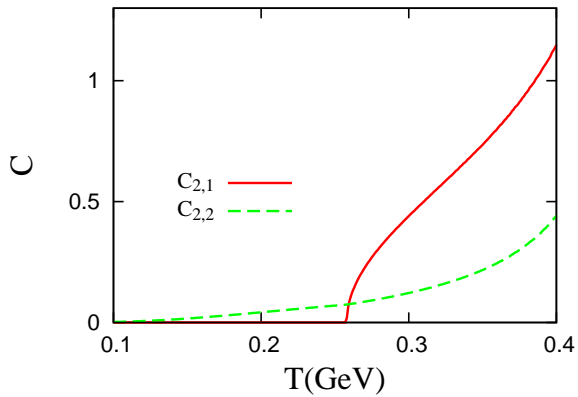}
\end{center}
\vspace{-10pt}
\caption{
$T$ dependence of the color state factors 
$C_{2,1}({\bf p}=0)$ (solid line) and $C_{2,2}({\bf p}=0)$ (dashed line) 
in the case of $\alpha =0$ and $(\theta_u,\theta_d)=(-\pi/2,\pi/2)$. 
}
\label{Fig_nc2_factor_c_z2b}
\end{figure}

In Fig. \ref{Fig_nc2_factor_c_z2b}, 
the color state factors $C_{2,1}({\bf p}=0)$ and 
$C_{2,2}({\bf p}=0)$ are drawn as a function of $T$. 
The one-quark state $C_{2,1}({\bf p}=0)$ vanishes below $T_c$ 
because of $\Phi=0$. Above $T_c$, on the contrary, the system is dominated 
by the one-color state, although the two-quark state $C_{2,2}$ remains there. 

The delay of the chiral restoration at higher $T$ 
can be understood as follows. 
Taking the flavor summation in \eqref{nc2_2} leads to 
\begin{align}
\Omega
&= -2 \sum_{c=1,2} \int \frac{d^3 p}{(2\pi)^3}
   \Bigl[ N_\mathrm{c} E_{f} \nonumber\\
&        + \frac{1}{\beta}\ln~ [1 + F_{2,1}({\bf p})e^{-\phi_c}+F_{2,2}({\bf p}) e^{-2i\phi_c}]
\notag\\
&        + \frac{1}{\beta}\ln~ [1 + F_{2,1}^*({\bf p})e^{i\phi_c}+F_{2,2}^*({\bf p}) e^{2i\phi_c}]
	      \Bigl]
	      \nonumber\\
& + U(\sigma , a_0) +{\cal U}(\Phi ,T), 
\label{nc2_6}
\end{align}
where 
\begin{eqnarray}
F_{2,1}({\bf p}) &=& e^{i\theta_u}e^{-\beta E_u}+e^{i\theta_d}e^{-\beta E_d} ,
\notag\\
F_{2,2}({\bf p}) &=& e^{i(\theta_u+\theta_d)}e^{-\beta (E_u+E_d)}. 
\label{F_cluster_nc2}
\end{eqnarray}
Since $\theta_u=\theta_1 $ and $\theta_d=\theta_1 +\pi$, 
Eq.~\eqref{F_cluster_nc2} is reduced to  
\begin{eqnarray}
F_{2,1}({\bf p}) &=& e^{i\theta_1}\left( z_{2,1}e^{-\beta E_u}+z_{2,2}e^{-\beta E_d}\right) ,
\notag\\
F_{2,2}({\bf p}) &=& -e^{2i\theta_1 }e^{-\beta (E_u+E_d)}, 
\label{F_cluster-2}
\end{eqnarray}
where $z_{2,1}=1$ and $z_{2,2}=-1$ are elements of the $\mathbb{Z}_2$ group. 
In the case of $(\theta_u,\theta_d)=(-\pi/2,\pi /2)$, the flavor symmetry is 
not broken, so that $E\equiv E_u=E_d$. In this situation, 
$F_{2,1}$ and $F_{2,2}$ are further reduced to 
\begin{eqnarray}
F_{2,1}({\bf p}) &=& -i\left( z_{2,1}+z_{2,2}\right)e^{-\beta E}=0
\notag\\
F_{2,2}({\bf p}) &=& e^{-2\beta E}. 
\label{F_cluster-2b}
\end{eqnarray}
The thermodynamic system thus has no $F_{2,1}$ but finite $F_{2,2}$. 
This means that u- and d-quarks are statistically in the same state. 
The chiral condensate $\sigma$ has weak $T$ dependence, since 
the two-quark state factor $F_{2,2}$ is strongly suppressed by the 
factor $\exp(-2\beta E)$. 
Eventually, the chiral restoration becomes much slower in the TBC model. 
This slow restoration is true also for the case of $N_c=3$ and 
$N_f=2$~\cite{Sakai_imiso}, although the $\mathbb{Z}_3$ symmetry 
is not exact in the case.

\begin{figure}[htbp]
\begin{center}
\hspace{-10pt}
\includegraphics[width=0.3\textwidth,angle=-90]{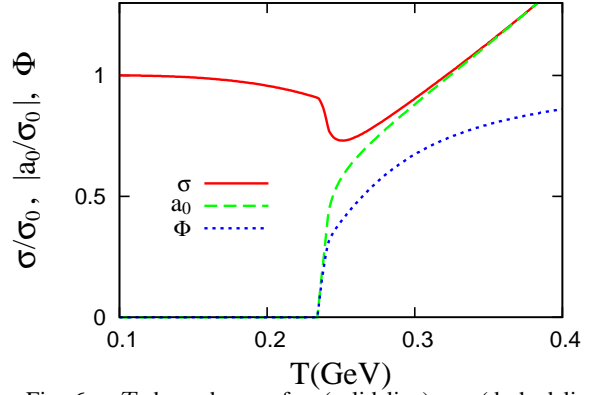}
\end{center}
\vspace{-10pt}
\caption{
$T$ dependence of $\sigma$ (solid line), $a_0$ (dashed line) 
and $\Phi$ (dotted line) in the case of $\alpha =0$ and 
$(\theta_u,\theta_d)=(0,\pi)$. 
$\sigma$ and $a_0$ are normalized by $\sigma_0$.  
Note that $\sigma <0$ and $a_0\ge 0$.  
}
\label{Fig_nc2_order}
\end{figure}

Next we analyze the case 
\bea
(\theta_u,\theta_d)=(0,\pi) 
\label{case-2-2}
\eea
corresponding to $\theta_1 =0$ in the TBC of \eqref{period}; 
see the left panel of Fig.~\ref{theta_f_nc2} for the twisted angles. 
Figure \ref{Fig_nc2_order} presents $T$ dependence of $a_0$ and $\Phi$. 
In this case, the flavor symmetry is explicitly broken by the TBC. 
The second-order deconfinement phase transition occurs at 
$T=T_c\approx 235$MeV. Below $T_c$, $a_0$ and $\Phi$ are zero, 
indicating that the flavor symmetry is restored by 
the color confinement. 
Above $T_c$, both $a_0$ and $\Phi$ become finite, indicating 
that the flavor and $\mathbb{Z}_2$ symmetries break 
simultaneously. 
At high $T$ where the flavor symmetry breaking is strong, $\sigma$ is 
getting large with respect to increasing $T$. 
This behavior is quite different from 
the corresponding behavior of $\sigma$ in the standard-PNJL model.

\begin{figure}[htbp]
\begin{center}
\hspace{-10pt}
\includegraphics[width=0.3\textwidth,angle=-90]{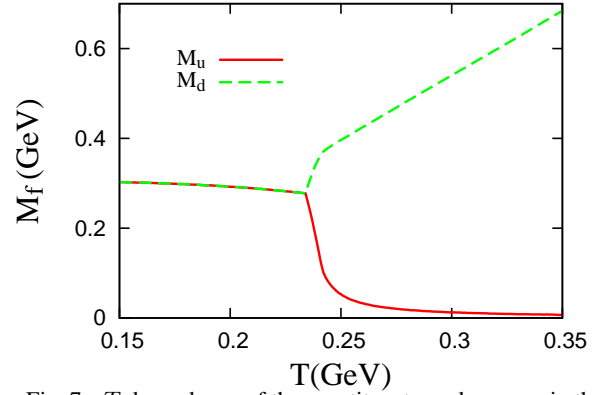}
\end{center}
\vspace{-10pt}
\caption{
$T$ dependence of the constituent quark masses in the case of 
$\alpha =0$ and $(\theta_u,\theta_d)=(0,\pi)$. 
The solid (dashed) line represents $u$ ($d$) quark mass.  
}
\label{Fig_nc2_mass}
\end{figure}

Figure \ref{Fig_nc2_mass} shows $T$ dependence of the constituent quark 
masses $M_u$ and $M_u$. The quark masses are degenerate below $T_c$, 
but above $T_c$ $d$-quark becomes heavier while $u$-quark does lighter. 
The mass splitting is a consequence of the flavor symmetry breaking.

\begin{figure}[htbp]
\begin{center}
\hspace{-10pt}
\includegraphics[width=0.3\textwidth,angle=-90]{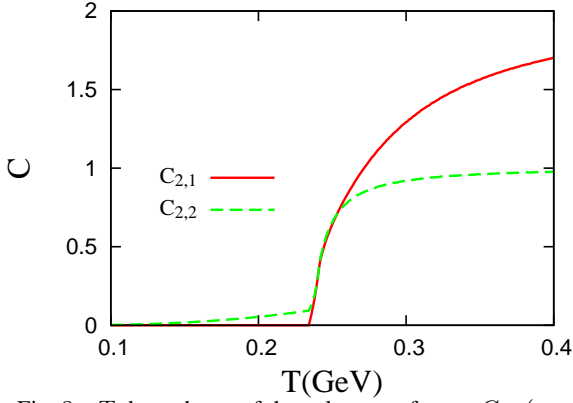}
\end{center}
\vspace{-10pt}
\caption{
$T$ dependence of the color state factors 
$C_{2,1}({\bf p}=0)$ (solid line) and $C_{2,2}({\bf p}=0)$ (dashed line) 
for $u$ quark in the case of 
$\alpha =0$ and $(\theta_u,\theta_d)=(0,\pi)$. 
}
\label{Fig_nc2_factor_c}
\end{figure}

In Fig. \ref{Fig_nc2_factor_c}, the color state factors 
$C_{2,1}({\bf p}=0)$ and $C_{2,2}({\bf p}=0)$  
are plotted for $u$-quark as a function of 
$T$. Below $T_c$, only the two-quark state $C_{2,2}$ remains. 
Above $T_c$, the system is dominated by the one-quark state 
$C_{2,1}$.

\begin{figure}[htbp]
\begin{center}
\hspace{-10pt}
\includegraphics[width=0.3\textwidth,angle=-90]{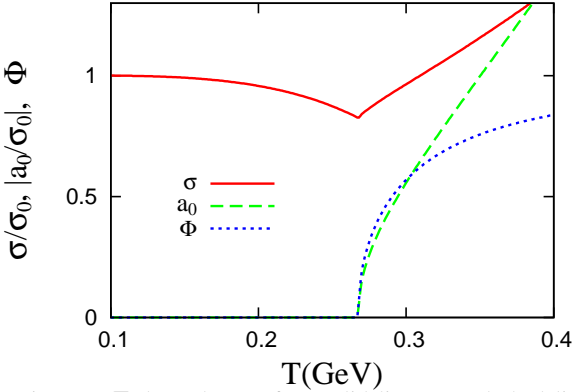}
\end{center}
\vspace{-10pt}
\caption{
$T$ dependence of $\sigma$ (solid line), $a_0$ (dashed line) and $\Phi$ 
(dotted line) in the case of $\alpha =0.2$ and $(\theta_u,\theta_d)=(0,\pi)$. 
$\sigma$ and $a_0$ are normalized by $\sigma_0$. 
Note that $\sigma <0$ and $a_0\ge 0$.  
}
\label{Fig_nc2_order_zeta_060}
\end{figure}

Figure \ref{Fig_nc2_order_zeta_060} shows $T$ dependence 
of $a_0$ and $\Phi$ in the case of $\alpha =0.2$. 
The $T$ dependence is similar to that in the case of $\alpha =0$, 
although $T_c\approx 265$~MeV in the former and $235$~MeV in the latter. 
Comparing Fig. \ref{Fig_nc2_order_zeta_060} with Fig. \ref{Fig_nc2_order}, 
one can see explicitly that $T_c$ becomes larger as $\alpha$ increases. 
The $U_{\rm A}(1)$ anomaly thus delays the spontaneous breaking of 
the $\mathbb{Z}_2$ symmetry and hence that of the flavor-symmetry breaking. 
 
\section{Case of $N_c=3$}
\label{Sec:NC3}
\subsection{Formalism}

The present formulation for $N_c=N_f=3$ is pararell to that for 
$N_c=N_f=2$ shown in Sec.~\ref{Formalism-color-2}. 
The PNJL Lagrangian with $N_c=N_f=3$ is  
\begin{align}
 {\cal L}  
=& \sum_f {\bar q}_f(\gamma_\nu D_\nu - \mu_f \gamma_4  + m_f )q_f  
\nonumber\\
  &- G_{\rm S} \sum_f \sum_{a=0}^{8} 
    [({\bar q}_f \lambda_a q_f )^2 +({\bar q }_fi\gamma_5 \lambda_a q_f )^2] 
\nonumber\\
 &+ G_{\rm D} \Bigl[\det_{ij} {\bar q}_i (1+\gamma_5) q_j 
           +\det_{ij} {\bar q}_i (1-\gamma_5) q_j \Bigr]
\nonumber\\
&+{\cal U}(\Phi [A],{\bar \Phi} [A],T) , 
\label{L_nc3}
\end{align} 
where $D_\nu=\partial_\nu + iA_\nu=\partial_\nu +i\delta_{\nu, 4}A_{4,a}{\tilde{\lambda}_a / 2}$ with the Gell-Mann matrices $\tilde{\lambda}_a$ 
in color space. 
In the interaction part, $\lambda_a~(a\neq 0)$ and $\lambda_0$ 
are the Gell-Mann matrices and the unit matrix in flavor space, respectively, 
and $G_{\rm S}$ and $G_{\rm D}$ are coupling constants 
of the scalar-type four-quark and the KMT determinant 
interaction~\cite{KMK,tHooft}, respectively, 
in which the determinant 
runs in flavor space. 
The KMT determinant interaction breaks the $U_\mathrm{A} (1)$ symmetry 
explicitly.

The Polyakov-loop $\Phi$ and its conjugate $\Phi ^*$ are determined by
\begin{align}
\Phi &= {1\over{3}}{\rm tr}_{\rm c}(L),
~~~~~\Phi^* ={1\over{3}}{\rm tr}_{\rm c}({\bar L}),
\label{Polyakov_nc3}
\end{align}
where $L  = \exp(i A_4/T)$ with 
$A_4/T={\rm diag}(\phi_r,\phi_g,\phi_b)$. Noting that $\phi_r+\phi_g+\phi_b=0$, 
one can obtain 
\begin{eqnarray}
\Phi &=&{1\over{3}}(e^{i\phi_r}+e^{i\phi_g}+e^{i\phi_b})
\notag\\
&=&{1\over{3}}(e^{i\phi_r}+e^{i\phi_g}+e^{-i(\phi_r+\phi_g)}), 
\notag\\
\Phi^* &=&{1\over{3}}(e^{-i\phi_r}+e^{-i\phi_g}+e^{-i\phi_b})
\notag\\
&=&{1\over{3}}(e^{-i\phi_r}+e^{-i\phi_g}+e^{i(\phi_r+\phi_g)}).
\label{Polyakov_explict_nc3}
\end{eqnarray}
We take the Polyakov potential of Ref.~\cite{Rossner}:
\begin{align}
&{\cal U} = T^4 \Bigl[-\frac{a(T)}{2} {\Phi}^*\Phi\notag\\
      &~~~~~+ b(T)\ln(1 - 6{\Phi\Phi^*}  + 4(\Phi^3+{\Phi^*}^3)
            - 3(\Phi\Phi^*)^2 )\Bigr] ,
            \label{eq:E13}\\
&a(T)   = a_0 + a_1\Bigl(\frac{T_0}{T}\Bigr)
                 + a_2\Bigl(\frac{T_0}{T}\Bigr)^2,~~~~
b(T)=b_3\Bigl(\frac{T_0}{T}\Bigr)^3 .
            \label{eq:E14}
\end{align}
Parameters of $\mathcal{U}$ are determined to reproduce LQCD data at finite $T$ in the pure gauge limit. 
The parameters except $T_0$ are summarized in Table \ref{table-para}.  
The Polyakov potential yields the first-order deconfinement phase transition 
at $T=T_0$ in the pure gauge theory~\cite{Boyd,Kaczmarek}. 
The original value of $T_0$ is $270$ MeV determined from the pure gauge 
LQCD data, but the PNJL model with this value of $T_0$ yields a larger 
value of the pseudocritical temperature $T_\mathrm{c}$ 
at zero chemical potential than $T_c\approx 160$~MeV predicted 
by full LQCD \cite{Borsanyi,Soeldner,Kanaya}. 
We then rescale $T_0$ to 195~MeV to reproduce 
$T_c=160$~MeV~\cite{Sasaki-T_Nf3}. 

\begin{table}[h]
\begin{center}
\begin{tabular}{llllll}
\hline \hline
~~~~~$a_0$~~~~~&~~~~~$a_1$~~~~~&~~~~~$a_2$~~~~~&~~~~~$b_3$~~~~~
\\
\hline
~~~~3.51 &~~~~-2.47 &~~~~15.2 &~~~~-1.75\\
\hline \hline
\end{tabular}
\caption{
Summary of the parameter set in the Polyakov-potential sector 
determined in Ref.~\cite{Rossner}. 
All parameters are dimensionless. 
}
\label{table-para}
\end{center}
\end{table}

Now we consider the flavor-dependent imaginary chemical potential 
$\mu_f=i\theta_fT$. The thermodynamic potential 
(per volume) based on the mean-field approximation is~\cite{Matsumoto}  
\begin{align}
\Omega
&= -2 \sum_{f=u,d,s} \sum_{c=r,g,b}\int \frac{d^3 p}{(2\pi)^3}
   \Bigl[ E_f \nonumber\\
&        + \frac{1}{\beta}\ln~ [1 + e^{i\phi_c}e^{i\theta_f} e^{-\beta E_f}]
\notag\\
&        + \frac{1}{\beta}\ln~ [1 + e^{-i\phi_c}e^{-i\theta_f}e^{-\beta E_f}]
\Bigl]
\nonumber\\
&+ U(\sigma_u,\sigma_d,\sigma_s) +{\cal U}(\Phi,\Phi^*,T) 
\label{PNJL-Omega_original}
\end{align}
with $\sigma_{f} \equiv \langle {\bar q}_f q_f \rangle$ and 
$E_f \equiv \sqrt{{\bf p}^3+{M_{f}}^2}$ for $f=u,d,s$, 
where the three-dimensional cutoff is taken 
for the momentum integration in the vacuum term~\cite{Matsumoto}. 
The dynamical quark masses $M_{f}$ are defined by 
\bea
M_{f}=m_{f}-4G_{\rm S}\sigma_{f}+  2G_{\rm D} \sigma_{f^\prime } \sigma_{f^{\prime\prime}} 
\eea
for $f \neq f^\prime$ and $f \neq f^{\prime\prime}$ and $f^{\prime\prime}\neq f^{\prime\prime\prime}$. 
The mesonic potential $U(\sigma_u,\sigma_d,\sigma_s)$ are obtained by
\bea
U(\sigma_u,\sigma_d,\sigma_s)=\sum_{f=u,d,s}  2 G_{\rm S} \sigma_{f}^2 
 - 4 G_{\rm D} \sigma_{u}\sigma_{d}\sigma_{s}. 
\label{Upotential}
\eea

For the 2+1 flavor system with $m_u=m_d\equiv m_l$, 
the PNJL model has five parameters 
($G_{\rm S}$, $G_{\rm D}$, $m_l$, $m_s$, $\Lambda$). 
A typical set is obtained in Ref.~\cite{Rehberg}. 
The parameter set is fitted to empirical values of 
$\eta'$-meson mass and $\pi$-meson mass and $\pi$-meson decay constant at vacuum. 
In the present paper, we set $m_s$ to $m_l$ in the parameter set 
of Ref.~\cite{Rehberg}, since we consider 
the three degenerate flavor system with $m_0 \equiv m_l=m_s$. 
The parameter set is shown in Table~\ref{Table_NJL}.

\begin{table}[h]
\begin{center}
\begin{tabular}{llllll}
\hline
~~$m_0(\rm MeV)$~~&~~$\Lambda(\rm MeV)$~~~&~~$G_{\rm S} \Lambda^2$
~~&~~$G_{\rm D} \Lambda^5$~~
\\
\hline
~~~~~~~~5.5 &~~~~~~~~602.3 &~~~~~~1.835 &~~~~~~12.36 &~~~~\\
\hline
\end{tabular}
\caption{
Summary of the parameter set in the NJL sector. 
All the parameters except $m_0$ are the same as in Ref.~\cite{Rehberg}. 
\label{Table_NJL}
}
\end{center}
\end{table}

Taking the color summation in \eqref{PNJL-Omega_original} leads to 
\begin{align}
\Omega
&= -2 \sum_{f=u,d,s} \int \frac{d^3 p}{(2\pi)^3}
   \Bigl[ 
   N_{\mathrm{c}} E_f 
   \nonumber\\
&        + \frac{1}{\beta}\ln [1 + C_{3,1}({\bf p}) e^{i\theta_f}
\notag\\
&+C_{3,2}({\bf p})e^{2i\theta_f}+ C_{3,3}({\bf p})e^{3i\theta_f}]
\notag\\
&        + \frac{1}{\beta}\ln [1 + 
C_{3,1}^*({\bf p})e^{-i\theta_f}e^{-\beta E_f} 
\notag\\
&+C_{3,2}^*({\bf p})e^{-2i\theta_f}+ C_{3,3}^*({\bf p})e^{-3i\theta_f}]
	      \Bigl]
	      \nonumber\\
&+ U(\sigma_u,\sigma_d,\sigma_s) +{\cal U}(\Phi,\Phi^*,T), 
\label{PNJL-Omega}
\end{align}
where 
\begin{eqnarray}
C_{3,1}({\bf p})&=&3\Phi e^{-\beta E_f}, 
\nonumber\\
C_{3,2}({\bf p})&=&3\Phi^* e^{-2\beta E_f}, 
\nonumber\\
C_{3,3}({\bf p})&=&e^{-3\beta E_f}. 
\label{factor_3_omega}
\end{eqnarray}

One can find that $\Omega$ has the RW periodicity, 
\begin{eqnarray}
\Omega (\theta_f)=\Omega (\theta_f+2k\pi /3),
\label{RW_c3}
\end{eqnarray}
making the ${\mathbb Z}_3$ transformation, 
\bea
\Phi  \to e^{-i{2\pi k/{3}}} \Phi, \quad
\Phi^{*} \to e^{i{2\pi k/{3}}}\Phi^{*},  
\label{Z3}
\eea
in $\Omega$. In the case of $(\theta_u,\theta_d,\theta_s)
=(\theta_1,\theta_1 +2\pi/3,\theta_1 +4\pi/3)$, $\Omega$ is invariant 
under the ${\mathbb Z}_3$ transformation, indicating that 
$\Omega$ possesses the $\mathbb{Z}_3$ symmety. 
When $\Phi$ vanishes, the color confinement ($C_{3,1}=C_{3,2}=0$) 
occurs and thereby $\Omega$ has the flavor symmetry ($E_u=E_d=E_s$) since 
the factors $e^{\pm 3i \theta_f}$ have no flavor dependence 
in \eqref{PNJL-Omega}. 
The flavor symmetry is thus preserved by 
the color confinement ($\Phi=0$) also for the case of $N_c=3$.

\subsection{Numerical results}

\begin{figure}[htbp]
\begin{center}
\hspace{-10pt}
\includegraphics[width=0.3\textwidth,angle=-90]{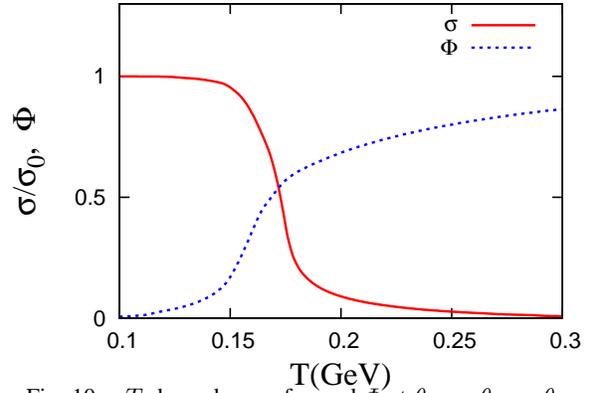}
\end{center}
\vspace{-10pt}
\caption{
$T$ dependence of $\sigma$ and $\Phi$ at $\theta_u=\theta_d=\theta_s=0$. 
$\sigma$ is normalized by $\sigma_0$.   
}
\label{nc3_mu0_order}
\end{figure}

First we consider the standard fermion boundary condition by setting 
$\theta_u=\theta_d=\theta_s=0$. 
In this case, the $\sigma_f$ are degenerate and hence 
$\sigma\equiv (\sigma_u+\sigma_d+\sigma_s)/3=\sigma_f$.  
Figure \ref{nc3_mu0_order} shows $T$ dependence of $\sigma$ and $\Phi$. 
Both the chiral restoration and the deconfinement transition are crossover, 
although the former transition is a bit slower 
than the latter~\cite{Sasaki-T_Nf3}.

\begin{figure}[htbp]
\begin{center}
\vspace{0.5cm}
\includegraphics[width=0.45\textwidth]{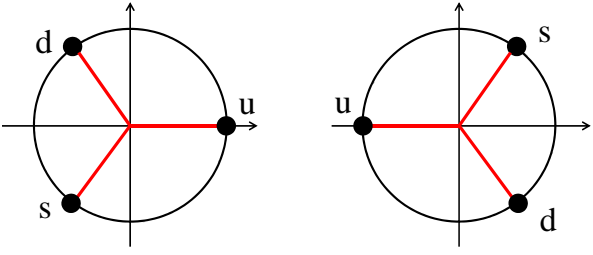}
\end{center}
\vspace{-10pt}
\caption{ Twisted factors $e^{i\theta_f}$ on a unit circle 
in the complex plane for the case of 
$\theta_1 =0$ (left) and $\theta_1 =-\pi$ (right). }
\label{theta_f_nc3}
\end{figure}

Next we consider the TBC model by taking two cases of 
$(\theta_u,\theta_d,\theta_s)=(0,2\pi /3,4\pi/3)$ and $(-\pi,-\pi /3,\pi/3)$ 
that 
correspond to the left and right panels in Fig. \ref{theta_f_nc3}, 
respectively. 
The charge conjugation yields the relation
\bea 
\Omega(\theta_u,\theta_d,\theta_s)=\Omega(-\theta_u,-\theta_d,-\theta_s)
=\Omega(\theta_u,\theta_s,\theta_d) 
\eea
for the two cases. 
Thus s-quark is symmetric with d-quark in these cases. 
Because of the $\mathbb{Z}_3$ symmetry, 
there are three $\mathbb{Z}_3$ vaccua when $\Phi \ne 0$. 
We then take the solution in which a phase $\phi$ of $\Phi$ lies in a 
range of $-\pi /3 \le \phi < \pi /3$. 
In the solution, $\Phi$ is found to be real. 

\begin{figure}[htbp]
\begin{center}
\hspace{-10pt}
\includegraphics[width=0.3\textwidth,angle=-90]{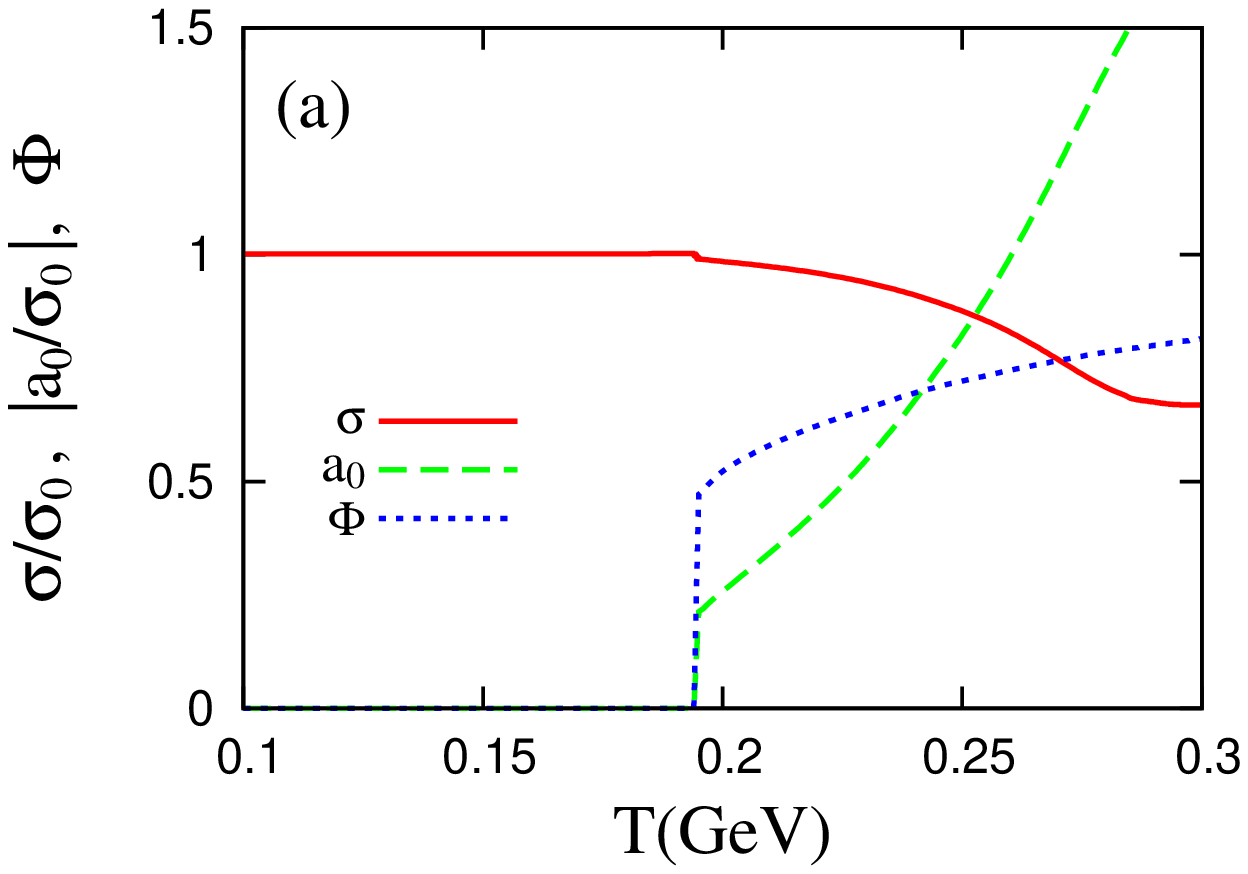}
\includegraphics[width=0.3\textwidth,angle=-90]{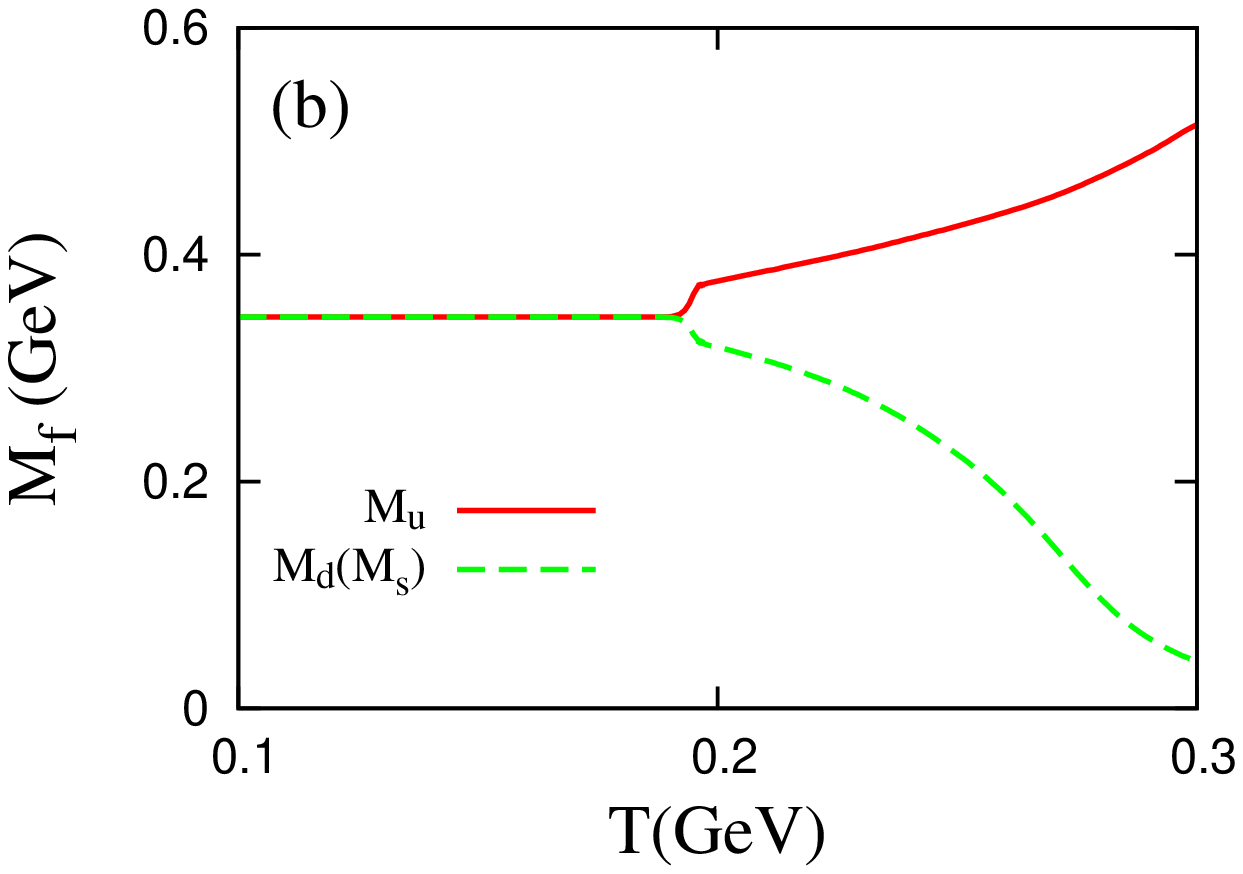}
\includegraphics[width=0.3\textwidth,angle=-90]{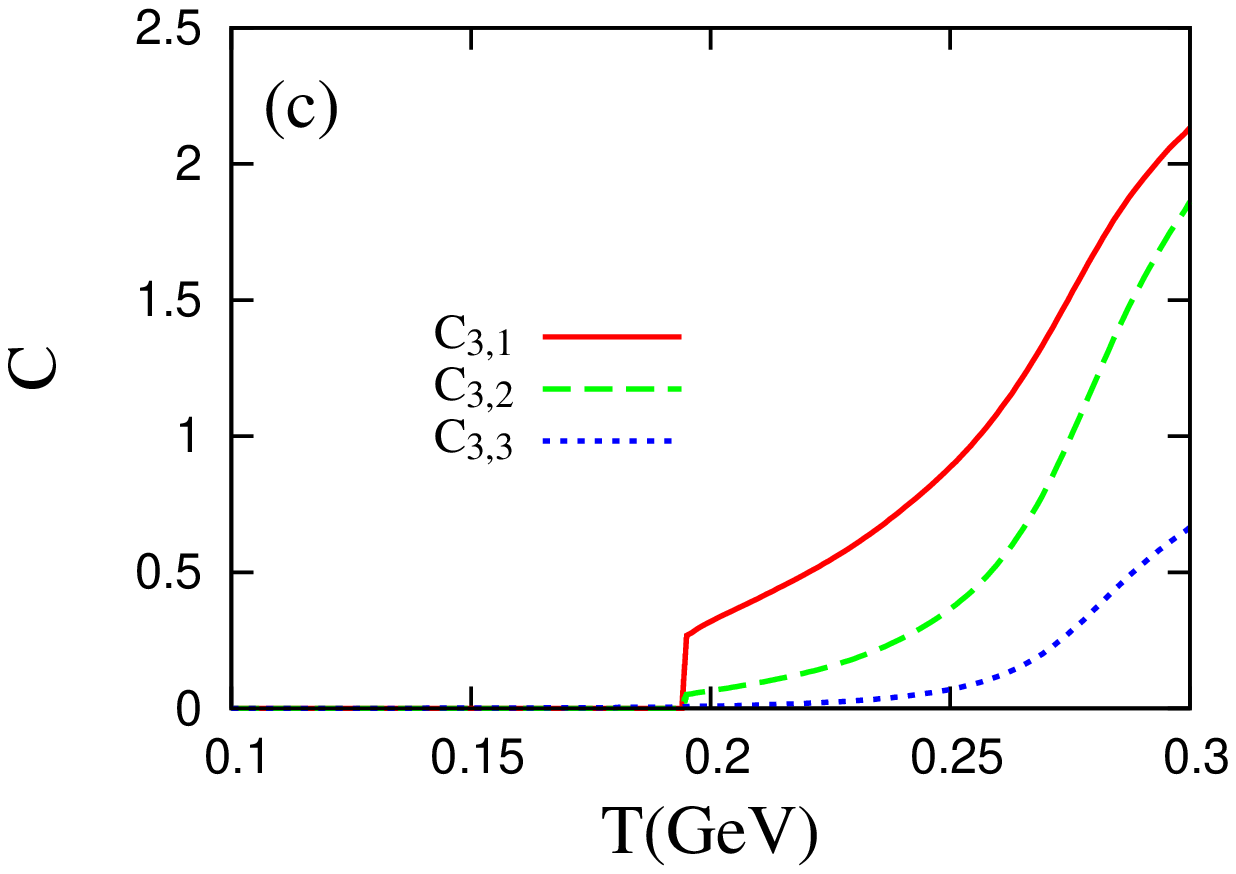}
\end{center}
\vspace{-10pt}
\caption{
$T$ dependence of (a) order parameters $\sigma$, $a_0$, $\Phi$, 
(b) constituent quark masses $M_f$, 
(c) color-state factors $C_{3,1}$, $C_{3,2}$, $C_{3,3}$ at ${\bf p}=0$ 
in the case of $(\theta_u,\theta_d,\theta_s)=(-\pi,-\pi /3,\pi/3)$. 
Here $\sigma$ and $a_0$ are normalized by $\sigma_0$. 
Note that $M_d=M_s$, $\sigma <0$ and $a_0\le 0$. 
}
\label{nc3_order_z3b}
\end{figure}

Figure \ref{nc3_order_z3b} shows $T$ dependence of several physical quantities 
in the case of $(\theta_u,\theta_d,\theta_s)=(-\pi,-\pi /3,\pi/3)$. 
The order parameters $\Phi$, $\sigma $ and 
$a_0\equiv  \sigma_u-\sigma_d=\sigma_u-\sigma_s$ are plotted in panel (a). 
The first-order deconfinement transition takes place 
at $T=T_c \approx 195$~MeV. 
Below $T_c$, $a_0$ and $\Phi$ are zero. 
The flavor symmetry is thus preserved by the color confinement. 
Above $T_c$, $a_0$ and $\Phi$ become finite, 
indicating that the flavor and $\mathbb{Z}_3$ symmetries break 
simultaneously.

For the constituent quark masses $M_f$ shown in panel (b), 
all the $M_f$ are degenerate below $T_c$. 
Above $T_c$, $M_u$ becomes heavier whereas two of the three, 
$M_d$ and $M_s$, are degenerate and becomes lighter. 
The increase of $M_u$ makes the chiral restoration slower. 
In panel (c), the absolute values of the color-state factors 
$C_{3,1}$, $C_{3,2}$ and $C_{3,3}$ are plotted at ${\bf p}=0$. 
Below $T_c$, $C_{3,3}$ is small but finite, whereas $C_{3,1}=C_{3,2}=0$. 
Above $T_c$, the system is dominated by the one-quark state $C_{3,1}$.

\begin{figure}[htbp]
\begin{center}
\hspace{-10pt}
\includegraphics[width=0.3\textwidth,angle=-90]{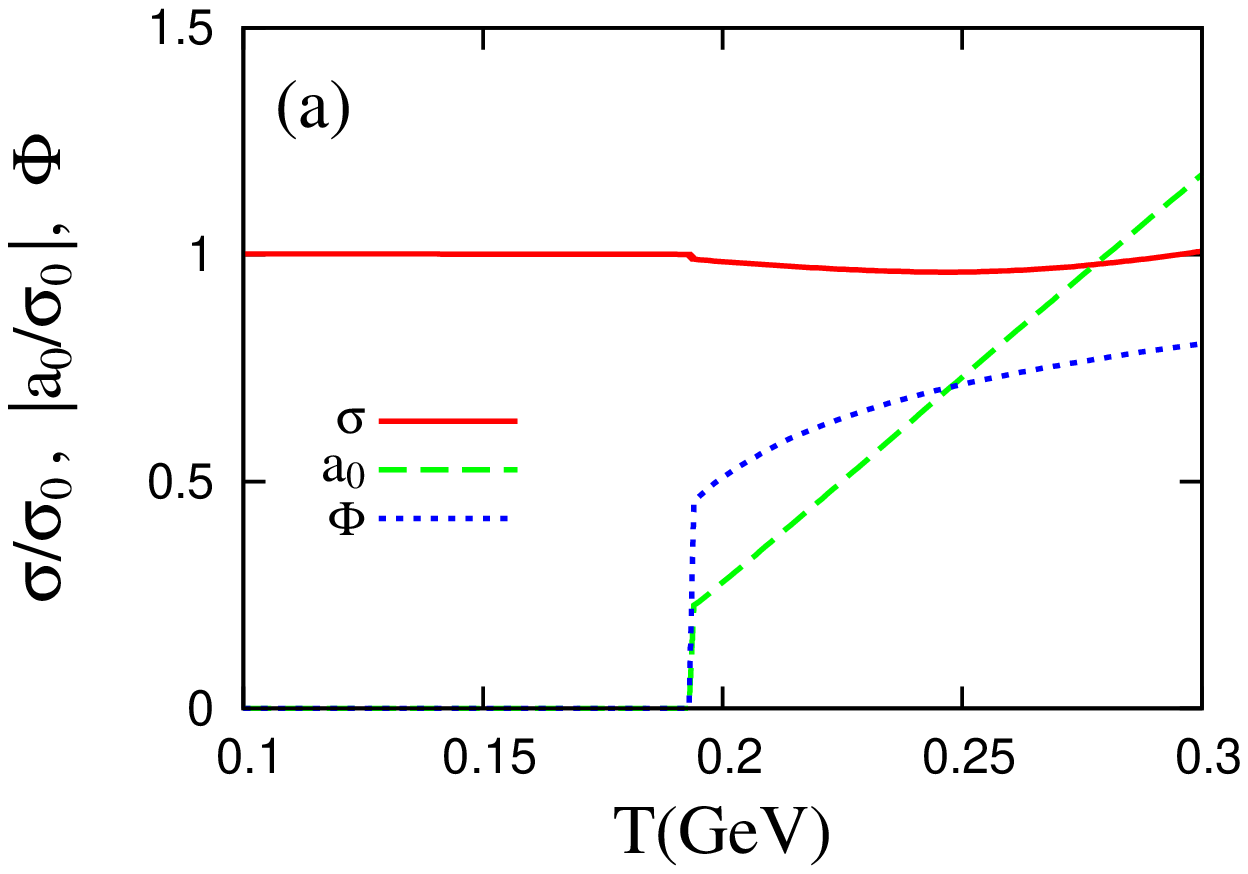}
\includegraphics[width=0.3\textwidth,angle=-90]{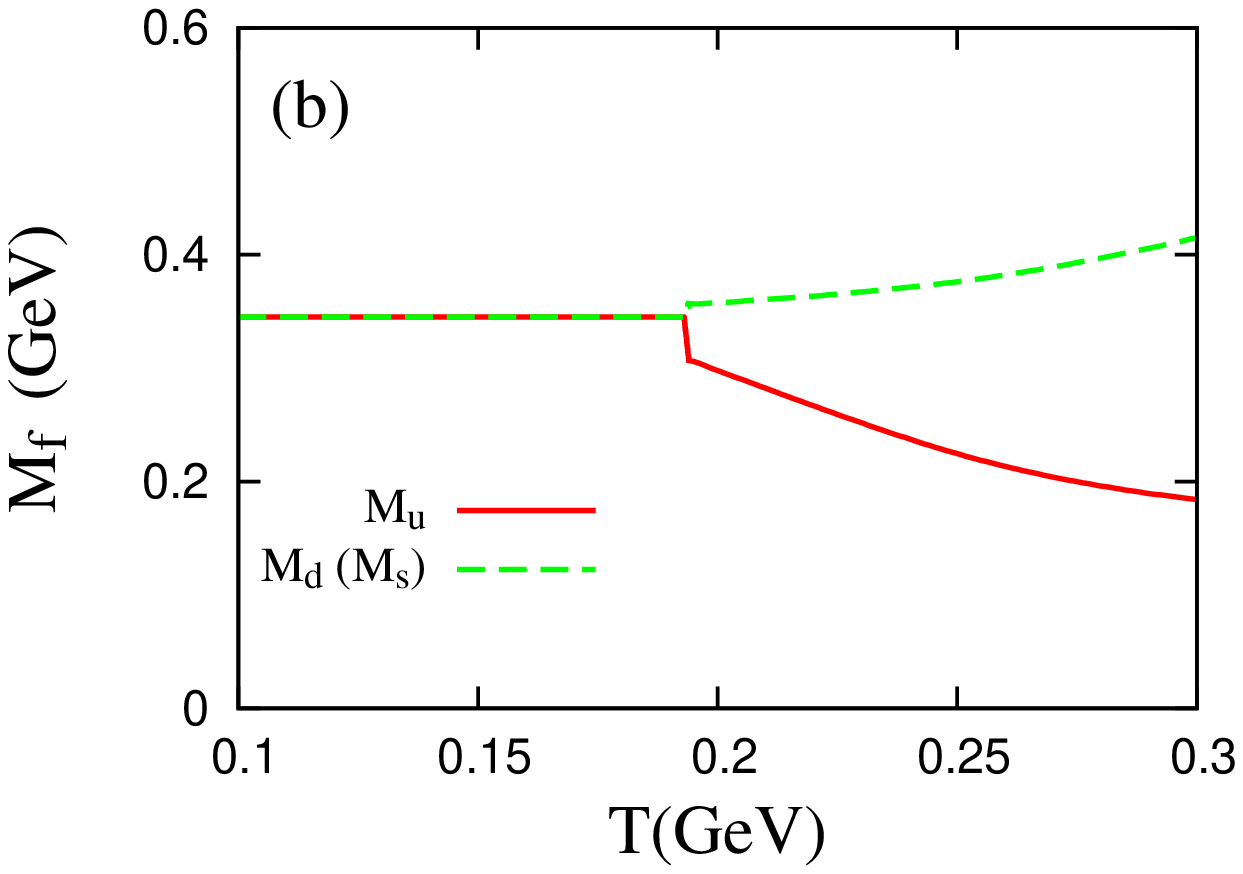}
\includegraphics[width=0.3\textwidth,angle=-90]{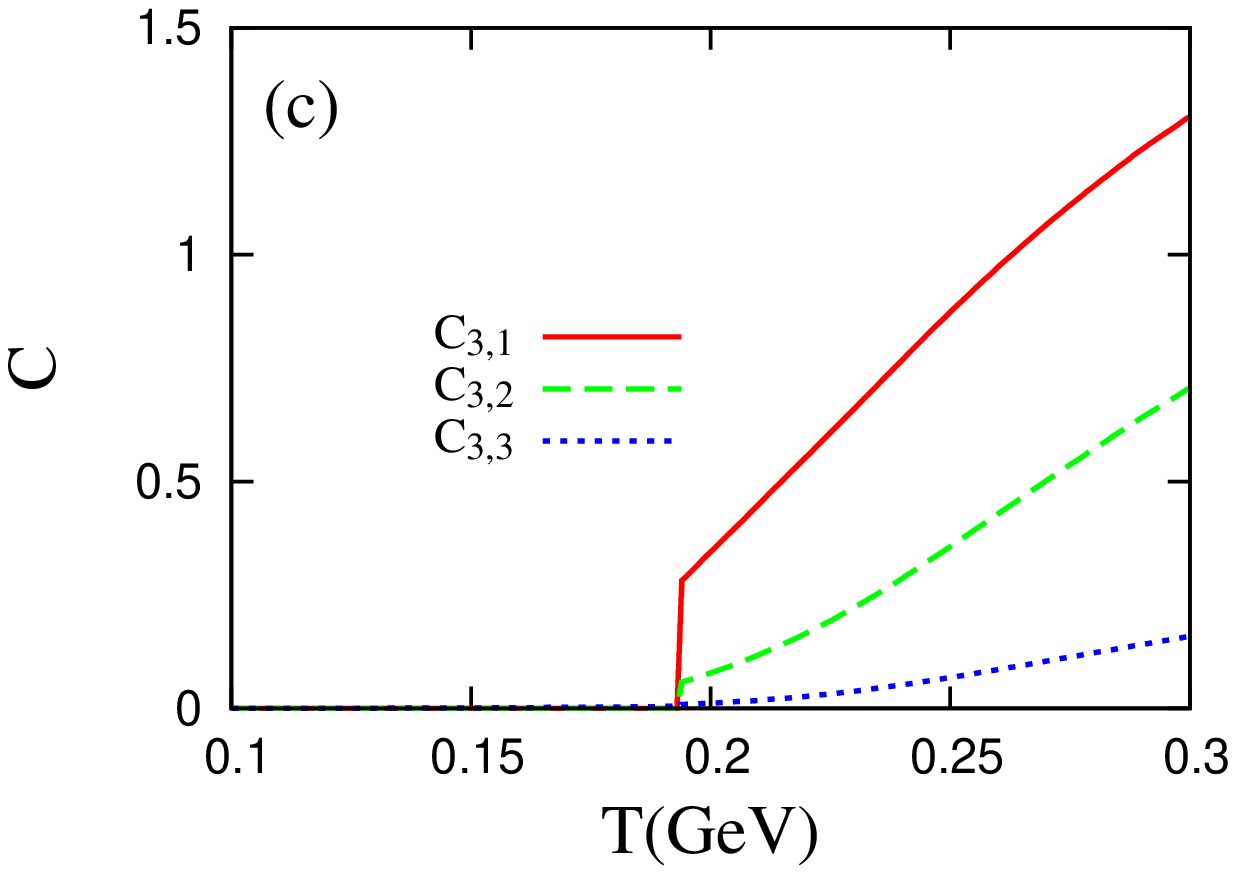}
\end{center}
\vspace{-10pt}
\caption{
$T$ dependence of (a) order parameters $\sigma$, $a_0$, $\Phi$, 
(b) constituent quark masses $M_f$, 
(c) color-state factors $C_{3,1}$, $C_{3,2}$, $C_{3,3}$ at ${\bf p}=0$ in 
the case of $(\theta_u,\theta_d,\theta_s)=(0,2\pi /3,4\pi/3)$. 
Here $\sigma$ and $a_0$ are normalized by $\sigma_0$. 
Note that $M_d=M_s$, $\sigma <0$ and $a_0\ge 0$.  
}
\label{nc3_order}
\end{figure}

Here the case of $(\theta_u,\theta_d,\theta_s)=(0,2\pi /3,4\pi/3)$ is 
considered briefly. 
As shown in Fig. \ref{nc3_order}, below $T_c\approx 195$MeV 
physical quantities have the same properties as those in the previous case. 
The difference between the two cases appears above $T_c$. 
Particularly for $M_f$, it is found that $M_d=M_s > M_u$ in the present case, 
while $M_d=M_s < M_u$ in the previous case. 
Thus both $d$- and $s$-quarks becomes heavier as 
$T$ increases from $T_c$. 
This property makes the chiral restoration even slower 
in the present case.

\subsection{Two extensions of the TBC model}
\label{Two extensions of the TBC model}

In this subsection, we extend the TBC model in two directions.

As the first extension, we use the entanglement PNJL (EPNJL) 
model~\cite{Sakai5,Sasaki-T_Nf3} instead of the PNJL model. 
A possible origin of the four-quark vertex $G_{\rm S}$ is a gluon exchange 
between quarks and its higher-order diagrams. 
If the gluon field $A_{\nu}$ has a vacuum expectation value 
$\langle A_{0} \rangle$, 
$A_{\nu}$ is coupled to $\langle A_{0} \rangle$  and hence 
to $\Phi$ through $L$~\cite{Kondo}. 
This effect allows $G_{\rm S}$ to depend on $\Phi$, namely 
$G_{\rm S}=G_{\rm S}(\Phi)$~\cite{Kondo}. 
It is expected that $\Phi$ dependence of $G_{\rm S}(\Phi )$ 
will be determined in future by accurate methods 
such as the exact renormalization group method~\cite{Braun,Kondo,Wetterich}. 
In this paper, however, we simply assume the following $G_{\rm S}(\Phi )$ 
by respecting the chiral symmetry, the charge-conjugation 
symmetry~\cite{Kouno} and the extended $\mathbb{Z}_3$ symmetry~\cite{Sakai}: 
\begin{eqnarray}
G_{\rm S}(\Phi)=G_{\rm S}[1-\alpha_1\Phi\Phi^*-\alpha_2(\Phi^3+\Phi^{*3})]. 
\label{entanglement-vertex}
\end{eqnarray}
The PNJL model with the entanglement vertex 
\eqref{entanglement-vertex} is called the EPNJL 
model~\cite{Sakai5,Sasaki-T_Nf3}. 
In principle, $G_{\rm D}$ can depend on $\Phi$, too. 
However, $\Phi$-dependence of $G_{\rm D}$ 
yields qualitatively the same effect on the phase diagram as that of 
$G_{\rm S}$~\cite{Sasaki-T_Nf3}. We then neglect 
$\Phi$-dependence of $G_{\rm D}$, following Ref.~\cite{Sasaki-T_Nf3}.

The parameters $\alpha_1$ and $\alpha_2$ in \eqref{entanglement-vertex} 
are so determined as to reproduce two results of LQCD at finite $T$. 
The first is a result of 2+1 flavor LQCD at $\mu=0$~\cite{YAoki} 
that the chiral transition 
is crossover at the physical point. The second is 
a result of degenerate three-flavor LQCD at $\theta=\pi$~\cite{FP2010} 
that the order of the RW endpoint is 
first-order for small and large quark masses but second-order 
for intermediate quark masses. 
The parameter set $(\alpha_1, \alpha_2)$ satisfying these conditions 
is located in the triangle region~\cite{Sasaki-T_Nf3}
\bea
\{-1.5\alpha_1+0.3 < \alpha_2 <-0.86\alpha_1+0.32,~\alpha_2 >0\}. 
\label{triangle}
\eea
As a typical example, we take $\alpha_1=0.25$ and $\alpha_2=0.1$,  
following Ref.~\cite{Sasaki-T_Nf3} and rescale $T_0$ 
to 150MeV~\cite{Sasaki-T_Nf3}.

\begin{figure}[htbp]
\begin{center}
\hspace{-10pt}
\includegraphics[width=0.3\textwidth,angle=-90]{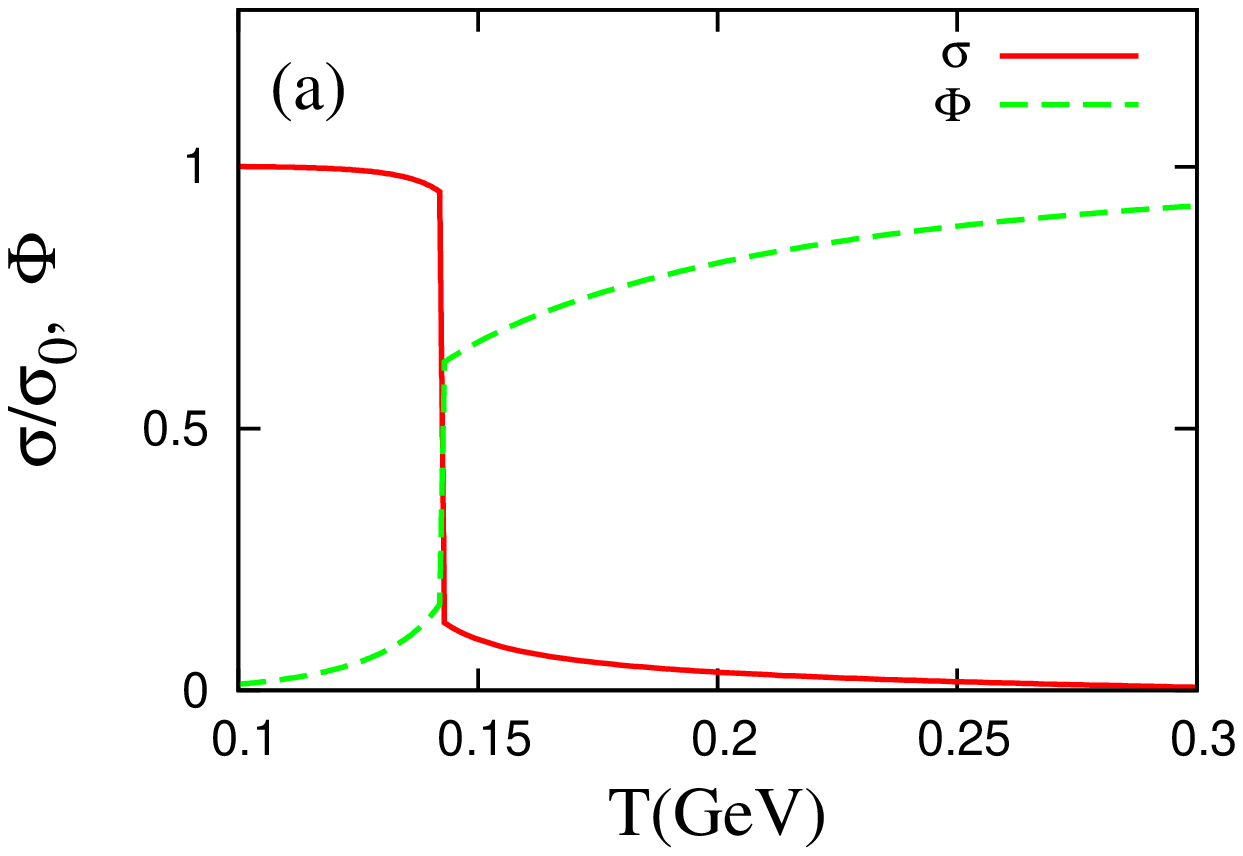}
\includegraphics[width=0.3\textwidth,angle=-90]{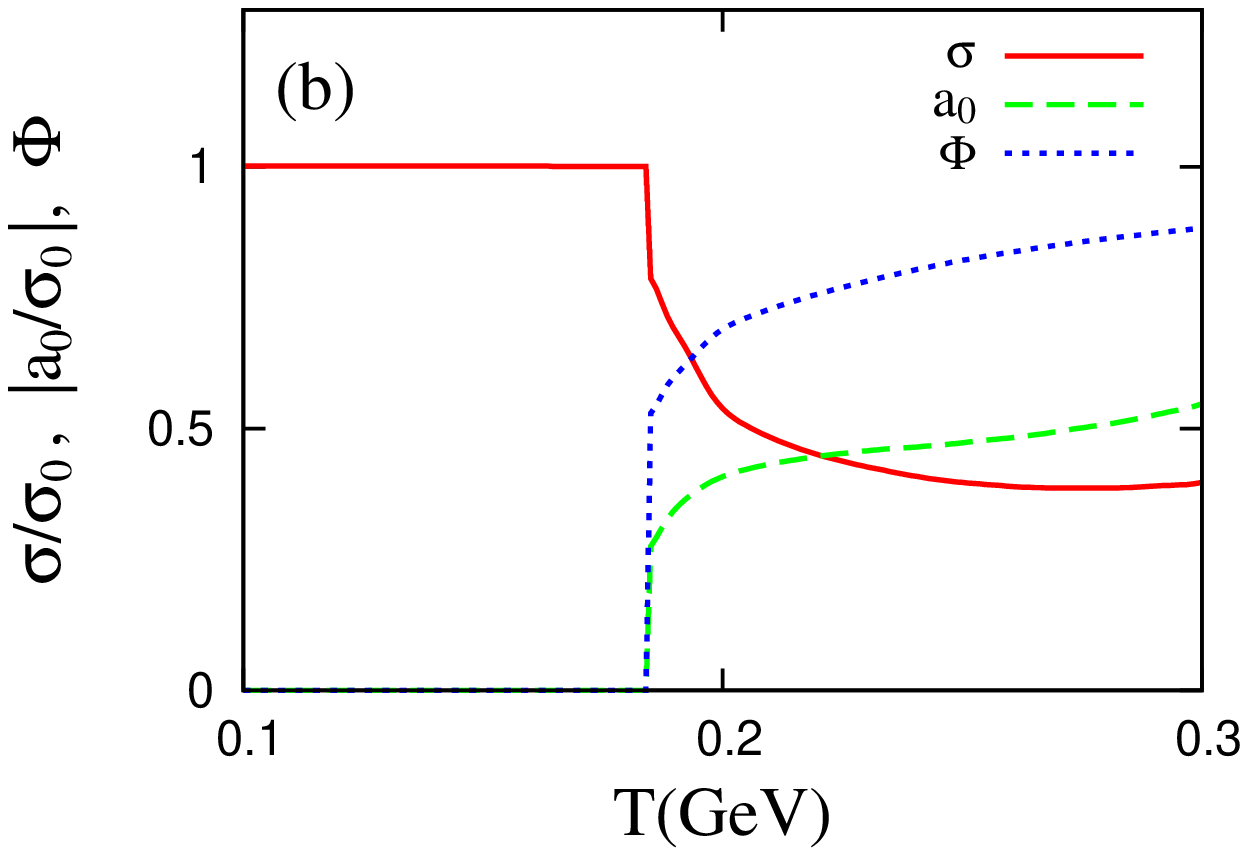}
\end{center}
\vspace{-10pt}
\caption{$T$ dependence of order parameters 
$\sigma$, $a_0$ and $\Phi$ 
calculated with the EPNJL model 
for (a) $(\theta_u,\theta_d,\theta_s)=(0,0,0)$ and 
(b) $(\theta_u,\theta_d,\theta_s)=(0,2\pi /3,4\pi/3$). 
Here $\sigma$ and $a_0$ are normalized by $\sigma_0$. 
Note that $a_0=0$ in panel (a) and $a_0\ge 0$ in panel (b), while $\sigma <0$ in both panels. 
}
\label{nc3_mu0_order_epnjl}
\end{figure}

Figure \ref{nc3_mu0_order_epnjl} shows $T$ dependence of 
$\sigma$, $a_0$ and $\Phi$ calculated with the EPNJL model 
for (a) $(\theta_u,\theta_d,\theta_s)=(0,0,0)$ and (b) 
$(\theta_u,\theta_d,\theta_s)=(0,2\pi /3,4\pi/3$). 
In panel (a), 
the chiral restoration and the deconfinement transition 
are first-order, 
because of the small current quark mass ($5.5$MeV) and 
the strong correlation between $\sigma_f$ and $\Phi$~\cite{Sasaki-T_Nf3}. 
In panel (b), the TBC model with the entanglement vertex yields similar $T$ 
dependence to the EPNJL model with the standard quark boundary condition 
for the chiral restoration and 
the deconfinement transition, although 
the flavor symmetry is broken above $T_c$.

As the second extension of the TBC model with $N_f=N_c$, 
one can consider the TBC model with $N_f=l N_c$ for any positive integer $l$. 
It is obvious that the TBC model with $N_f=l N_c$ has the 
$\mathbb{Z}_{N_c}$ symmetry, if the twisted angles $\theta_f$ are 
properly ordered; for example,  
\bea
\theta_f=\theta_1 +2 \pi(f-1)/N_f, 
\label{extended-TBC}
\eea 
or 
\bea
\theta_f=\theta_1 +2\pi(f-1)/N_c, 
\label{extended-TBC-2}
\eea 
for $f=1, 2, \cdots, N_f$. 

\begin{figure}[htbp]
\begin{center}
\vspace{0.5cm}
\includegraphics[width=0.48\textwidth]{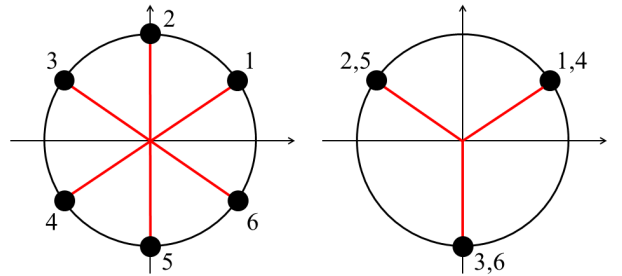}
\end{center}
\vspace{-10pt}
\caption{Twisted factors $e^{i\theta_f}$ on a unit circle 
in the complex plane in the case of $N_c=3$, $N_f=6$ and $\theta_1 =\pi/6$.
In the left and right panels, the $e^{i\theta_f}$ are obtained by 
\eqref{extended-TBC} and \eqref{extended-TBC-2}, respectively. 
}
\label{Nf6}
\end{figure}

Let us consider the case of $N_c=3$, $N_f=6$ and $\theta_1=\pi/6$. 
In Fig.~\ref{Nf6}, the left and right panels show 
the twisted angles defined by \eqref{extended-TBC} and 
\eqref{extended-TBC-2}, respectively. 
Here we take the right-panel case as an example. 
The thermodynamic potential $\Omega$ has the same form 
as \eqref{PNJL-Omega}, except the flavor summation is taken from $f=1$ to 6. 
It is straightforward to show that $\Omega$ has the RW periodicity 
and the ${\mathbb Z}_3$ symmetry.

We take the same parameter set as in the case of $N_f=N_c=3$, except 
$G_{\rm s}$ is taken as 
$G_{\rm S}=G_{\rm S,3}-G_{\rm D,3}\sigma_{f}(0)/2=2.226~{\rm GeV}^2$, 
where 
$G_{\rm S,3}$ and $G_{\rm D,3}$ mean $G_{\rm S}$ 
and $G_{\rm D}$ in the case of $N_c=N_f=3$, respectively, and 
$\sigma_{f}(0)$ stands for $\sigma_f$ at $T=0$ and $\theta_f=0$ 
in the case of $N_c=N_f=3$. 
We keep the Polyakov potential $\cal{U}$ of \eqref{eq:E14}, but 
neglect the KMT determinant interaction just for simplicity. 

Figure \ref{Fig_nc2_order} presents $T$ dependence of $\sigma_f$ and $\Phi$ 
for the right-panel case of Fig.~\ref{Nf6}. 
Below $T_c \approx 190$~MeV, the flavor symmetry ($\sigma_1=\sigma_2=\cdots 
=\sigma_6$) is preserved by the color confinement ($\Phi=0$). 
Above $T_c$, the flavor and $\mathbb{Z}_{3}$ symmetries break simultaneously. 
The flavor symmetry breaking is partial because 
$f=1$ is symmetric with $f=4$, $f=3$ with $f=6$, and $f=2$ with $f=5$.  
As a consequence of this property, the $\sigma_f$ split into three doublets.

\begin{figure}[htbp]
\begin{center}
\hspace{-10pt}
\includegraphics[width=0.3\textwidth,angle=-90]{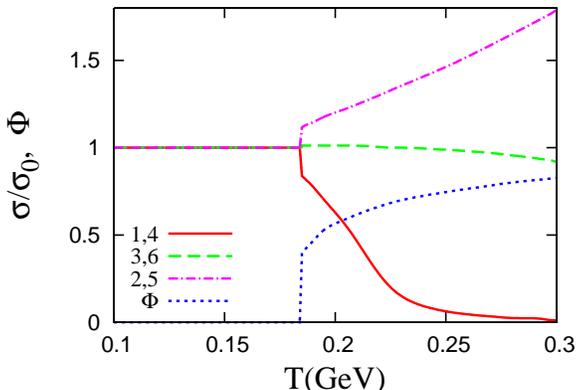}
\end{center}
\vspace{-10pt}
\caption{
$T$ dependence of $\sigma_f$ and $\Phi$ in the right-panel case of 
Fig.~\ref{Nf6}. 
The solid, dashed and dot-dashed lines represent 
$\sigma_1=\sigma_4$, $\sigma_3=\sigma_6$ and $\sigma_2=\sigma_5$, 
respectively, whereas the dotted line corresponds to $\Phi$.
}
\label{Fig_nc3_nf6}
\end{figure}


\section{Summary}
\label{Summary}
We have proposed a QCD-like theory with the $\mathbb{Z}_{N_c}$ symmetry. 
The QCD-like theory is constructed by imposing 
the flavor-dependent twisted boundary condition \eqref{period} on 
the SU($N_c$) gauge theory 
with $N_c$ degenerate flavor quarks. 
Dynamics of the QCD-like theory has been studied 
by imposing the TBC on the PNJL model. 
The TBC model has the $\mathbb{Z}_{N_c}$ symmetry and hence the 
Polyakov loop becomes an exact order parameter of the deconfinement transition. 
The TBC model is a good model to investigate the mechanism of 
color confinement.

For both cases of $N_f=N_c=2$ and 3, 
the Polyakov loop is zero up to some temperature $T_c$, but becomes 
finite above $T_c$. The $\mathbb{Z}_{N_c}$ symmetry is thus preserved 
below $T_c$, but spontaneously broken above $T_c$. 
Below $T_c$, the color confinement preserves the flavor symmetry. 
Above $T_c$, meanwhile, the flavor symmetry is broken explicitly by 
the TBC. 
The flavor-symmetry breaking makes the chiral restoration slower, but 
the entanglement interaction between $\sigma$ and $\Phi$ makes 
the restoration faster. The entanglement interaction thus suppresses 
the flavor symmetry breaking. 
In the standard-PNJL model with degenerate flavor quarks, 
$\Phi$ becomes finite but small at $T$ lower than the pseudo-critical 
temperature, while the flavor symmetry is preserved. 
Dynamics of the TBC model is thus 
similar to that of the standard-PNJL model below $T_c$. 
The similarity is relatively worse above $T_c$, but 
it is improved by the entanglement interaction. 
One can then expect that 
QCD with the approximate $\mathbb{Z}_{3}$ 
symmetry is similar to the QCD-like theory with the $\mathbb{Z}_{3}$ 
symmetry and hence that the $\mathbb{Z}_{3}$ symmetry is 
a good approximate concept in QCD, 
even if the current quark mass is small.

The model prediction mentioned above can be tested with LQCD, 
since LQCD with the TBC has no sign problem. 
The QCD-like theory is useful to understand the mechanism of 
color confinement, since the $\mathbb{Z}_{N_c}$ symmetry is exact. 
For example, it is quite interesting 
to see $T$ dependence of the potential between $q$ and $\bar{q}$ 
in LQCD with the TBC.

\noindent
\begin{acknowledgments}
The authors thank A. Nakamura, T. Saito, K. Nagata and K. Kashiwa for useful discussions. 
H.K. also thanks M. Imachi, H. Yoneyama, H. Aoki and M. Tachibana for useful discussions. 
T.S and Y.S. are supported by JSPS. 
The calculation was partially carried out on SX-8 at Research Center for Nuclear Physics, Osaka University. 
\end{acknowledgments}


\end{document}